\documentclass[iop]{emulateapj}

\usepackage{hyperref}
\usepackage{xcolor}
\usepackage{url}
\usepackage{amssymb}
\usepackage{physics}
\hypersetup{
    colorlinks,
    linkcolor={red!50!black},
    citecolor={blue!50!black},
    urlcolor={black} 
}
\usepackage{mathrsfs}
\usepackage{nicefrac}

\usepackage{lineno}
\shorttitle{Primary Velocity and Orbital Phase Effects on Planetary Detectability}
\shortauthors{Buzard et al.}

\begin{document}

\title{Primary Velocity and Orbital Phase Effects on Planetary Detectability from Small Epoch Number Data Sets}

\author{Cam Buzard\altaffilmark{1}, Stefan Pelletier\altaffilmark{2}, Danielle Piskorz\altaffilmark{3}, Bj{\"o}rn Benneke\altaffilmark{2}, Geoffrey A. Blake\altaffilmark{1,3}}

\altaffiltext{1}{Division of Chemistry and Chemical Engineering, California Institute of Technology, Pasadena, CA 91125}
\altaffiltext{2}{Institute for research on exoplanets, Universit{\'e} de Montr{\'e}al, Montreal, QC}
\altaffiltext{3}{Division of Geological and Planetary Sciences, California Institute of Technology, Pasadena, CA 91125}

\begin{abstract}
Cross correlation analyses of high resolution spectroscopic data have recently shown great success in directly detecting planetary signals and enabling the characterization of their atmospheres. One such technique aims to observe a system at multiple epochs and combine the measured planetary radial velocities from each epoch into a measurement of the planetary Keplerian orbital velocity $K_p$, constituting a direct detection of the planetary signal. Recent work has shown that in few-epoch ($\sim$5) data sets, unintended structure can arise at a high level, obscuring the planetary detection. In this work, we look to simulations to examine whether there are ways to reduce this structured noise in few-epoch data sets by careful planning of observations. The choice of observation date allows observers to select the primary (stellar) velocity --- through a set systemic velocity and chosen barycentric velocity --- and the planetary orbital phase, and so we focus on the effects of these two parameters. We find that epochs taken when the primary velocity is near zero, and the stellar lines remain relatively fixed to the telluric rest-frame, greatly reduce the level of structured noise and allow for much stronger planetary detections, on average more than twice the significance of detections made with epochs using randomly selected primary velocities. Following these results, we recommend that observers looking to build up high-resolution multi-epoch data sets target nights when their system has a near-zero primary velocity.     
\end{abstract}

\keywords{techniques: spectroscopic --- planets and satellites: atmospheres}

\section{Introduction}

As thousands of exoplanets are being discovered through indirect methods such as transit and radial velocity surveys, astronomers have begun to consider how to follow-up on these detections and measure the planets' atmospheric properties, especially the presence and relative abundances of molecular species, the atmospheric pressure/temperature profiles, and the nature of winds and planetary rotation. High resolution cross correlation spectroscopy has been introduced as an effective technique to directly detect planets' thermal emission and begin to characterize their atmospheres \citep[e.g., ][]{Brogi2012,Birkby2013,lockwood}. High resolution cross correlation spectroscopy works by allowing researchers to disentangle planetary and stellar radial velocities. By fitting the planetary radial velocities with an equation for the orbital motion, observers can constrain the amplitude of that motion, called the planetary Keplerian orbital velocity, $K_p$. With prior knowledge of the stellar mass and the stellar Keplerian orbital velocity $K_*$ from optical radial velocity measurements, we can constrain the true mass and orbital inclination of the planet. Further, planetary models with different assumptions about various atmospheric properties can be cross correlated against the data and the resulting strength of the planetary detection can be used to understand the true nature of the planetary atmosphere.  

Two approaches have been applied to constrain $K_p$ from the planetary radial velocities. In one, observers target a system at times when the line-of-sight planetary acceleration is largest, e.g. near inferior/superior conjunction \citep[e.g., ][]{Snellen2010,Brogi2012,Brogi2013,Birkby2017,Guilluy2019}. Then, by observing for many ($\sim$5-7) hours in one stretch, they can watch the planetary signal shift systematically with respect to the fixed stellar and telluric reference frames, as its line-of-sight velocity changes. The Keplerian velocity $K_p$ is measured through a fit to this changing planetary radial velocity. While this technique has been very effective for hot Jupiters, it requires a significant change in the line-of-sight orbital velocity (tens of km/s) over the course of a single continuous observing sequence. This will preclude its application to longer period planets, including those is the nearest M star habitable zones. Another approach, first introduced by \citet{lockwood}, limits wall-clock observing times to $\sim$2-3 hours to obtain measurements that do not allow the planetary signal to cross detector pixels. After building up a data set of several such measurements around the planet's orbit, the data can be fit to constrain $K_p$. We will call this approach the ``multi-epoch" approach.

\citet{Buzard2020} used the multi-epoch approach to detect the thermal emission from the hot Jupiter, HD187123b. In that work, we introduced a simulation frame that could account for a portion of the structured noise that arose in the $K_p$ detection space. With these simulations, we showed how beneficial many-epoch ($\sim$20) data sets are over few-epoch ($\sim$5) data sets, even if they share the same total S/N. This was because in the few-epoch simulations structured noise resulting from unwanted correlation between the planetary spectral template and the observed stellar signal was well in excess of the shot noise readily obtainable with Keck on bright stars. Large number epoch data sets afford more variation between the stellar and planetary signals that works to beat down this source of structured noise. However, it can be difficult to build up to such large data sets with highly over-subscribed telescopes such as Keck, especially with the current generation of high resolution echelle spectrographs that have modest instantaneous spectral grasp in the thermal infrared, such that significant integration times are needed per epoch. In this work, we investigate whether there is a way to provide more efficient detections with few-epoch data sets by carefully selecting which nights we choose to observe. To do so, we consider the effects of the primary (stellar) velocities and orbital phases at each epoch. 

At any given observation time, the stellar velocity in a system will be determined by the systemic velocity, the barycentric velocity from the component of the Earth's orbital motion in the direction of the system, and the radial velocity caused by the planetary tug on the star. The current NIRSPEC does not have the velocity precision necessary to resolve the radial velocity caused by a planet; while NIRSPEC $L$ band velocity precision is $\sim3.2$ km/s, the stellar RVs caused by even the massive hot Jupiters are $\lesssim 0.1$--$0.2$ km/s. Assuming a constant systemic velocity, then, the barycentric velocity is variable and can be chosen by when the observing night is scheduled. We will consider the primary velocity at a given observation time $t_{\mathrm{obs}}$, 

\begin{equation}
    v_{\mathrm{pri}}(t_{\mathrm{obs}}) = v_{\mathrm{sys}} - v_{\mathrm{bary}}(t_{\mathrm{obs}}). \label{Equation:primaryvelocity}
\end{equation}

The planetary velocity is similarly comprised of a dynamical radial velocity (which, unlike the stellar RV, is large enough to be resolved by NIRSPEC), the systemic velocity, and the barycentric velocity. The magnitude of the planetary gravitational radial velocity signature is $K_p$, and depends on the planetary and stellar masses, and the orbital inclination, semi-major axis, and eccentricity. The orbital phase, $M$, of the planet will determine the magnitude of the planetary radial velocity given $K_p$, and will therefore determine the planetary velocity relative to the stellar velocity. If the radial velocity parameters have been determined through optical, stellar RV measurements, $M$ can be calculated at any time through the equation,

\begin{equation}
    M(t_{\mathrm{obs}}) = \frac{(t_{\mathrm{obs}} - T_0) \mod P}{P}, 
\end{equation}
where $T_0$ is the time of inferior conjunction and $P$ is the orbital period. As a function of $t_{\mathrm{obs}}$, $M$ can also be chosen with careful observation scheduling. With these two parameters, and assuming a circular orbit, the secondary, or planetary, radial velocity can be described as

\begin{equation}
    v_{sec}(t_{obs}) = K_p\sin(2\pi M(t_{obs})) + v_{pri}(t_{obs}). \label{Equation:secondaryvelocity}
\end{equation}

Since the primary velocity ($v_{\mathrm{pri}}$) and orbital phase ($M$) can both be selected by the choice of observing nights, we set out to understand how different combinations of primary velocity and orbital phase epochs affect the detectability of planetary Keplerian orbital velocities, $K_p$, for a modest, and readily obtainable, 5 epoch data set. 

The rest of this work is organized as follows. In Section~\ref{Section:methods}, we describe the planetary spectral models used in these simulations, how the simulations are generated, and how they are analyzed. In Section~\ref{Section:simulations}, we consider the effectiveness of different groupings of orbital phases and primary velocities. We examine whether the magnitude of $K_p$ affects the results of these primary velocity/orbital phase simulations in Section~\ref{Section:KpMagnitude}. In Section~\ref{Section:ComparetoDataSection}, we attempt to see whether a combination of NIRSPEC data epochs agree with these simulation results. In Section~\ref{Section:EpochNumbers}, we consider the primary velocity and orbital phase effects on larger data sets. Finally, we discuss some implications of these results in Section~\ref{discuss} and conclude in Section~\ref{conclude}.

\section{Methods} \label{Section:methods}
\subsection{Spectral Models Used} \label{Section:spectralmodels}
For these simulations, we used a spectral model generated from the PHOENIX stellar spectral modeling framework \citep{Husser2013}. We interpolated the effective temperature, metallicity, and surface gravity to those of the sun-like star HD187123 ($T_{\mathrm{eff}} = 5815$ K, [Fe/H] = 0.121, and $\log(g) = 4.359$; \citet{valenti2005}). 

Our planetary thermal emission model was generated from the SCARLET framework \citep{Benneke2012, Benneke2013, benneke, Benneke2019a, Benneke2019} at $R = 250,000$. This framework computes equilibrium atmospheric chemistry and temperature structure assuming a cloud-free atmosphere with a solar elemental composition, efficient heat redistribution, and an internal heat flux of 75 K. We assume a solar metallicity and C/O ratio. The SCARLET model framework includes molecular opacities of H$_2$O, CH$_4$, HCN, CO, CO$_2$, NH$_3$, and TiO from the ExoMol database \citep{Tennyson2012, Tennyson2020}, molecular opacities of O$_2$, O$_3$, OH, C$_2$H$_2$, C$_2$H$_4$, C$_2$H$_6$, H$_2$O$_2$, and HO$_2$ (HITRAN database by \citet{Rothman2009}), alkali metal absorptions (VALD database by \citet{Piskunov1995}), H$_2$ broadening \citep{Burrows2003}, and collision-induced broadening from H$_2$/H$_2$ and H$_2$/He collisions \citep{Borysow2002}. The atmosphere does not have an inverted thermal structure in regions close to the molecular photosphere.

\subsection{Generation of Simulated Data} \label{Section:generatingsimulations}
In this work, we generated simulated data sets following the framework introduced by \citet{Buzard2020}. In short, the stellar and planetary models are scaled by assumed stellar and planetary radii squared and shifted to velocities determined from Equations~\ref{Equation:primaryvelocity} and \ref{Equation:secondaryvelocity}. Next, the stellar spectrum is interpolated onto the planetary wavelength axis, and the two models are added. The stellar continuum is removed with a third order polynomial fit to the combined spectrum from 2.8 to 4 $\mu$m in wavenumber space. The spectrum is broadened with a Gaussian kernal fit to real NIRSPEC data. Finally, the spectrum is interpolated onto a NIRSPEC data wavelength axis, saturated telluric pixels from the data (where tellurics absorb more than about 40\% of the flux) are masked, and Gaussian noise is added. The masking of saturated telluric removes about 40\% of the data. Non-saturated tellurics are assumed to be perfectly corrected.

For these simulations, we assume a 1 $R_{\mathrm{Jup}}$ planet and a 1 $R_{\sun}$ star. Unless otherwise stated, these simulations approximate post-upgrade $L$ band data. The upgraded NIRSPEC instrument was first available in early 2019 \citep{martin2018}. Across the $L$ band, it doubled the number of pixels per order (from 1024 to 2048), increased the number of usable orders on the chip (from 4 to 6), and nearly doubled the spectral resolution (from $\sim$25,000 to $\sim$40,000). The Gaussian kernals used to broaden the simulated data and wavelength axes with their corresponding locations of saturated telluric pixels were taken from the 2019 Apr 3 and 2019 Apr 8 NIRSPEC data of HD187123 presented in \citet{Buzard2020}. Each epoch has six orders, covering wavelengths of approximately 2.9331--2.9887, 3.0496--3.1076, 3.1758--3.2364, 3.3132--3.3765, 3.4631--3.5292, and 3.6349--3.6962 $\mu$m. The average instrumental resolution is about 41,000. We applied the average S/N per pixel from the 2019 Apr 3 and 8 data of 2860 to each epoch, which resulted in a total S/N per pixel of 6390 for the 5 epoch simulations.

\subsection{Analysis of Simulated Data}

The simulated data sets are analyzed analogously to the data presented in past multi-epoch detection works, e.g., \citet{Piskorz2018}, \citet{Buzard2020}. A two-dimensional cross correlation, TODCOR as described in \citet{zuckertodcor}, is used to measure the stellar and planetary velocities at each epoch. Segments of data (e.g., orders and pieces of orders after saturated telluric pixels are masked) are cross correlated separately and converted to log likelihood functions in order to be combined. In this work, we use the \citet{zucker2003} $\log(L)$ approach to convert cross correlations to log likelihoods; the ``Zucker $\log(L)$" approach is described and differentiated from the ``Zucker ML" approach in \citet{Buzard2020}. This approach converts cross correlations to log likelihoods as
\begin{equation}
    \log (L) = -\frac{n}{2}\log(1-R^2),
\end{equation}
where $R$ is the two-dimensional cross correlation and $n$ is the number of pixels in the data segment.

Once the two-dimensional cross correlation of each epoch is converted to a two-dimensional (stellar and planetary velocity shifts) log likelihood surface, the planetary log likelihood cut is taken from the measured stellar velocity. The measured stellar velocity is always consistent with the expected stellar velocity from Equation~\ref{Equation:primaryvelocity}. The planetary log likelihood curves at each epoch are converted from $v_{sec}$ to $K_p$ space by Equation~\ref{Equation:secondaryvelocity}, and added. This planetary log likelihood vs. $K_p$ curve is calculated from $-150 \leq K_p \leq 150$ km/s.

\section{Primary Velocity Simulations} \label{Section:simulations}

To study how primary velocities and orbital phases affect planetary detectability, we generate sets of simulated data with different combinations of primary velocities and orbital phases. These data sets all consider five epochs and have $K_p$ set at 75 km/s. For these simulated data sets, we allow $v_{pri}$ to range from -30 to 30 km/s, the rough maximum variation given by the Earth's orbital velocity, $v_{bary}$. We create five different groupings of primary velocities: (1) a most blue-shifted $v_{pri}$ sample, in which the primary velocities at all five epochs are pulled from a uniform distribution from -30 to -28 km/s; (2) an even $v_{pri}$ sample, in which the five primary velocities are evenly spaced from -30 km/s to 30 km/s; (3) a most red-shifted $v_{pri}$ sample, in which the five epochs are pulled from a uniform distribution from 28 to 30 km/s; (4) a near-zero $v_{pri}$ sample in which the five primary velocities are pulled from a uniform distribution from -2 to 2 km/s; and (5) a random $v_{pri}$ sample. For the evenly spaced $v_{pri}$ sample, the five epochs are pulled from uniform distributions covering: -30 to -29 km/s, -16 to -14 km/s, -1 to 1 km/s, 14 to 16 km/s, and 29 to 30 km/s. These slight variations in the most blue-shifted, most red-shifted, even, and near-zero primary velocity groups better resemble actual observations that could be scheduled than if all five large $v_{pri}$ epochs had exactly 30 km/s, for example. For the randomly sampled primary velocity group, we choose the Earth's orbital position from the uniform distribution from 0 to $2\pi$. These positions are then converted to barycentric velocities assuming the maximum barycentric velocity is 30 km/s. This results in a bimodal barycentric velocity distribution that is relatively uniform through the central velocities and increases significantly towards $\pm30$ km/s. We consider a systemic velocity of 0 km/s, so the resulting random primary velocity distribution has the same shape as the barycentric velocity distribution ($v_{pri} = -v_{bary}$). If the systemic velocity were non-zero, the primary velocity distribution would be shifted and the probability would increase towards its maximum ($v_{sys} - \mathrm{min}(v_{bary})$) and minimum ($v_{sys} - \mathrm{max}(v_{bary})$) values. We discuss how this may affect the results of the random primary velocity simulations in Section~\ref{discuss}. Realistically, systems are not observable from the Earth for the full year. A pull from half of the Earth's orbit (e.g., $v_{\mathrm{bary}} = 30\cos(0)$ to $30\cos(\pi)$) results in the same random primary velocity probability distribution, so we use this moving forward. We contemplate further effects of target accessibility in Section~\ref{discuss}.

We split combinations of orbital phases, $M$, up into three groups: (1) all five epochs near conjunction, (2) all five epochs near quadrature, and (3) five epochs evenly spaced around the orbit. The five near-conjunction epochs are pulled randomly from the uniform distributions, $0 \pm 0.02$ (inferior conjunction) and $0.5 \pm 0.02$ (superior conjunction), and the quadrature epochs are pulled from the uniform distributions, $0.25 \pm 0.02$ and $0.75 \pm 0.02$. The evenly spaced $M$ epochs have one epoch pulled from similarly wide uniform distributions centered on each of 0.05, 0.25, 0.45, 0.65, and 0.85. This is only one example of an evenly distributed set of orbital phases, and we expand the analysis to include other combinations of orbital phases in Section~\ref{Section:randomorbphases}.

\begin{figure*}
    \centering
    \noindent\includegraphics[width=42pc]{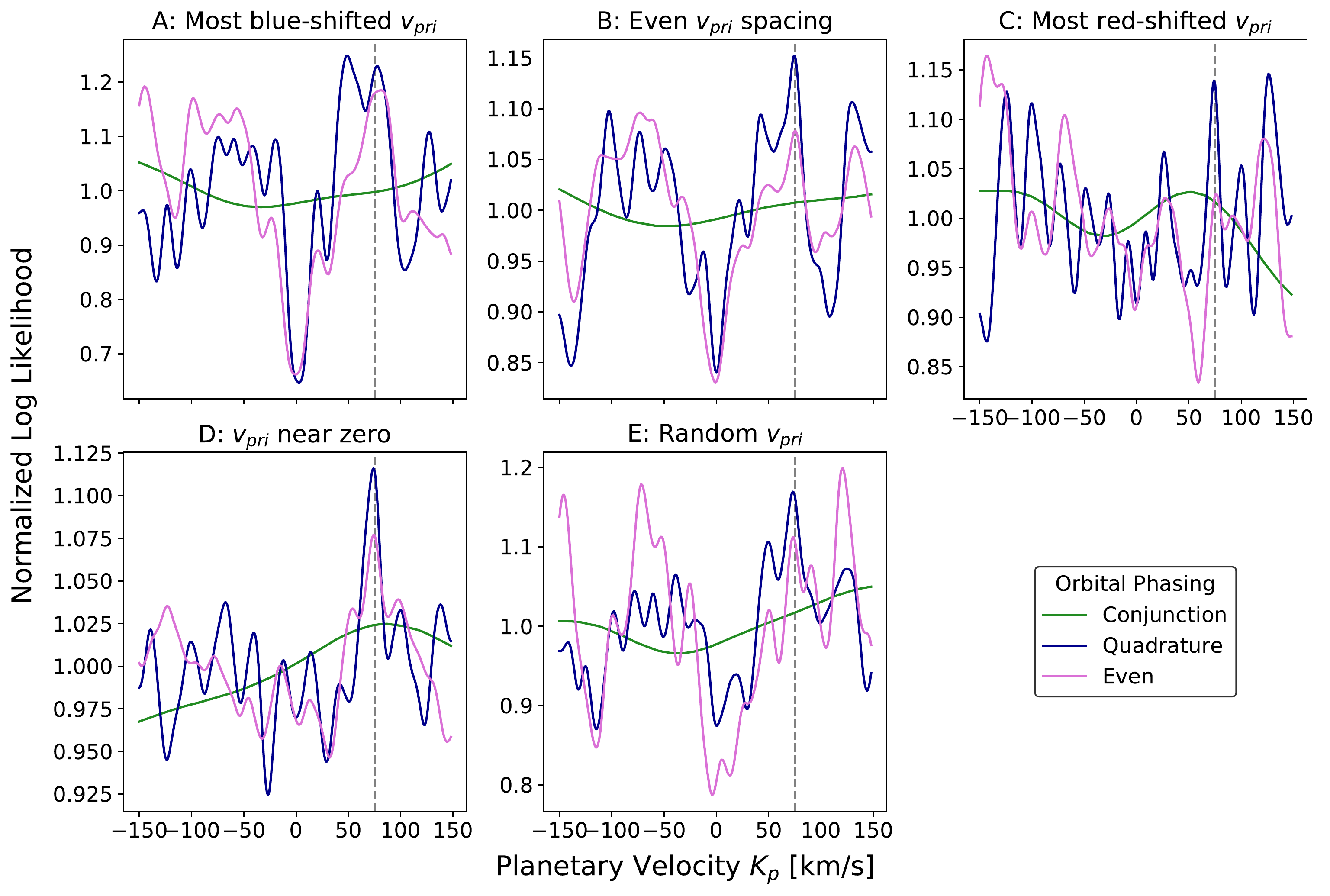}
    \caption{ Normalized log likelihood functions vs. $K_p$ of 5 epoch simulations with different combinations of primary velocities and orbital phases. The panels show five different groupings of primary velocities: most blue-shifted, evenly spaced, most red-shifted, near-zero, and random. The colors represent different combinations of orbital phases with near conjunction epochs in green, quadrature epochs in dark blue, and epochs evenly spaced around the orbit in light purple. The simulations with primary velocities near zero in each of the five epochs show less structured noise than do simulations with any other combination of primary velocities.   }
    \label{simulationsfigure}
\end{figure*}

Figure~\ref{simulationsfigure} shows the results of these combinations of primary velocities and orbital phases, with the five primary velocity groups taking up different subplots, and the three orbital phase groups shown in different colors. For each primary velocity subplot, near quadrature orbital phases are shown in dark blue, evenly spaced orbital phases in light purple, and near conjunction phases in green.

Several notable results stand out. First, there is very little structure in any of the simulations that considered all five epochs near conjunction. These simulations show no convincing detections of $K_p$. This trend makes sense because when the planet is near either inferior or superior conjunction, it will have little to no line-of-sight velocity difference from its star. Regardless of what $K_p$ is, the planet line-of-sight velocity at conjunction will simply be equal to the primary velocity. Conjunction epochs on their own are not useful for constraining the Keplerian orbital velocity through the technique that aims to measure stationary planetary velocities at multiple epochs. This is in contrast to cross correlation techniques that aim to measure changing planetary velocities \citep[e.g., ][]{Snellen2010, Brogi2012}; they actually prefer conjunction epochs, during which the planetary acceleration is the largest. It is also useful to note that these techniques that target changing planetary velocities would also require higher spectral resolution than the technique that targets stationary planetary velocities. 

The simulations with evenly spaced orbital phases and orbital phases near quadrature do have a peak at $K_p$ in all of the primary velocity groups, but there is often large off-peak structure at the same magnitude, if not larger, than the true peak. Both \citet{Buzard2020} and \citet{Finnerty} found that at the S/N used in these simulations ($>2500$ per pixel per epoch), shot noise has very little effect on the log likelihood surface, and the off-peak structure is due instead to non-random, structured noise. This structured noise is caused by correlation between the planetary spectral model template and the stellar features in the simulated data. \citet{Finnerty} removed this structure in their simulations by subtracting a stellar-only log likelihood curve, which comes from simulated data generated with no planetary signal and then cross correlated in the same two-dimensional way with a stellar and a planetary spectral model. This approach can nearly eliminate all of the off-peak structure in simulations, but it would not be as effective on data due to a variety of factors including mismatches between the real and model stellar and planetary spectra and imperfect removal of tellurics from the data. We therefore do not try to remove this off-peak structure. Instead, we look for combinations of primary velocities and orbital phases that can reduce it by design. 

Interestingly, we see in Figure~\ref{simulationsfigure} that the simulations with primary velocities around 0 seem to show stronger peaks at $K_p$ relative to the noise than for the other primary velocity groups. This appears to be true for both the evenly spaced orbital phases and the near quadrature orbital phases. While these simulations consider a planetary spectral model without a thermal inversion, simulations generated and analyzed with a planetary model with an inversion showed a similar trend in that near-zero primary velocity epochs produced the strongest detections.

\subsection{Random Orbital Phases} \label{Section:randomorbphases}

In order to investigate whether this trend that epochs taken when the primary velocity of the system is near zero provide stronger detections is more broadly true, we generated five-epoch data sets within each of the five primary velocity sampling groups, but with orbital phases pulled from a uniform distribution from 0 to 1, i.e. the full orbit. These data sets are likely more representative of real data sets that could be obtained from systems of interest too. While the barycentric velocity changes over the course of an Earth year, the orbital phase changes on the time frame of the planet's year. For hot Jupiters, the orbital phase changes significantly from night to night, making it difficult to obtain multiple epochs with the same orbital phase (especially when trying to schedule nights which will provide useful epochs for multiple targets). The primary velocity, on the other hand, will be approximately the same on a monthly timescale. So, simulations with set primary velocities and randomly picked orbital phases might be a good approximation to data sets that could be easily obtained.

\begin{figure*}
    \centering
    \noindent\includegraphics[width=42pc]{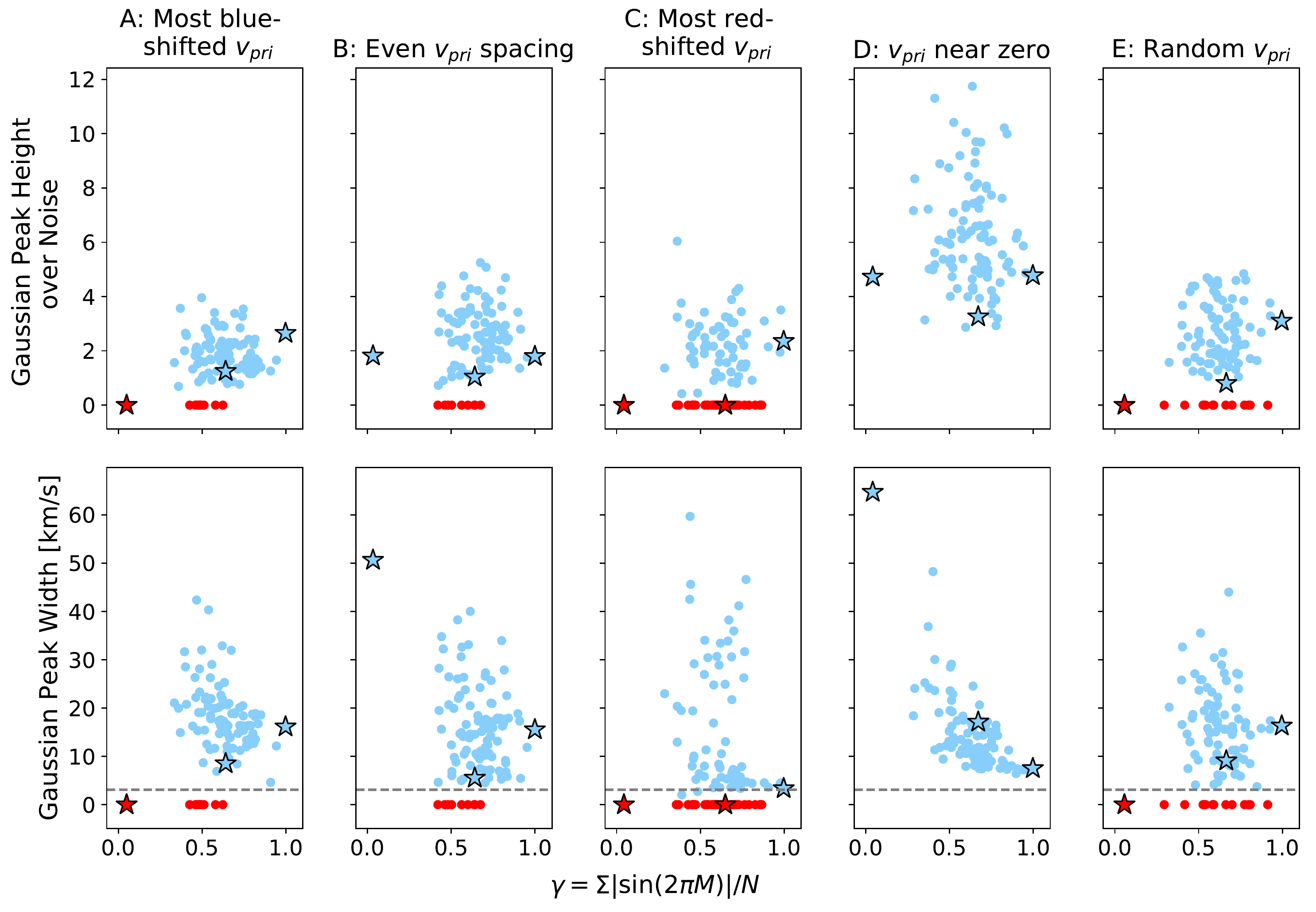}
    \caption{ Results of Gaussian fits to 100 simulations with randomly selected orbitals phases in each of five primary velocity groups. The top panels plot the heights of the Gaussian fits over the noise level and the bottom panels plot the Gaussian widths. Light blue points represent simulations with detectable planetary peaks and red points at 0 represent simulations with non-detections, in which the Gaussian mean was more than $1\sigma$ from the set $K_p$ of 75 km/s. The stars are the Gaussian parameters fit to the simulations in Figure~\ref{simulationsfigure}. The horizontal, gray, dashed line in the Gaussian width plots shows the approximate velocity precision of post-upgrade NIRSPEC, 3.1 km/s. The planet detections made with simulated data sets with 5 near-zero primary velocity epochs are significantly stronger than those made with any other combination of primary velocity epochs. For near-zero primary velocity simulations, epochs near quadrature put much stronger constraints on $K_p$ than do epochs far from quadrature.       }
    \label{guassianfitstorandomMfigure}
\end{figure*}

We generate 100 sets of five epoch simulations with randomly chosen orbital phases for each of the five primary velocity groups. We define a parameter, $\gamma$, to quantify each combination of orbital phases, as follows.
\begin{equation}
    \gamma = \frac{1}{N} \Sigma_i |\sin(2\pi M_i)|,      \label{Equation:gamma}
\end{equation}
where $N$ is the number of epochs. For epochs at quadrature, $\sin(2\pi M) = \pm 1$, and for epochs at conjunction, $\sin (2\pi M) = 0$. Defined this way, $\gamma = 0$ if all of the epochs are at conjunction and $\gamma = 1$ if all of the epochs are at quadrature. %For these 5 epoch simulations, $\gamma$ will range from 0 to 5.   

After analyzing each of the 500 simulations, we attempt to fit Gaussians to the resulting normalized log likelihood curves. We first normalize the curves by subtracting the mean of the curve from -150 to 0 km/s. We choose to normalize by the mean of this range because the way that we define $K_p$ enforces that it must be a positive value, meaning that the log likelihood curve at negative values of $K_p$ must be completely due to structured noise, and not to any real planetary signal. Because the simulated data is generated with a $K_p$ of 75 km/s, we fit Gaussians with an initial mean of 75 km/s, $\sigma$ of 10 km/s, and height equal to the normalized log likelihood value where $K_p = 75$ km/s.  

The results of these Gaussian fits are shown in Figure~\ref{guassianfitstorandomMfigure}. In the top row, we plot the height of each Gaussian fit over the standard deviation of the curve from -150 to 0 km/s. We use this as a means to show how much stronger the true planetary peak is than the structured noise. In the bottom row are the standard deviations of the Gaussian fits. We plot a blue point for each simulation for which a Gaussian could be fit within $1\sigma$ of the true value of $K_p$, 75 km/s, and a red point at 0 for both of the Gaussian parameters when it could not. The stars are the Gaussian fits to the simulations shown in Figure~\ref{simulationsfigure}.   

These results confirm that primary velocity near 0 km/s will generally allow for stronger detections of the planetary signal and more confident measurements of $K_p$. The planetary peak was detected in all 100 of the near-zero primary velocity cases, but only 91 of the even primary velocity cases, 64 of the most red-shifted primary velocity cases, 90 of the most blue-shifted primary velocity cases, and 86 of the random primary velocity cases. Further, the heights of the Gaussians fits to the near-zero primary velocity cases relative to the noise are much larger, on average, than for any of the other primary velocity cases. 

Interestingly, we see no obvious relationship between the $\gamma$ for a simulation and its peak height over the noise for any of the primary velocity groups. We suspect that at the larger values of $\gamma$, near quadrature, the planetary peak becomes resolved, leading to a larger height, but the noise structure also becomes narrower and of larger amplitude, so the increase in the peak height and noise level balance each other out. With the lack of dependence on $\gamma$, we can consider the mean and standard deviation of the peak heights over noise. Considering only those simulations in which the planetary peak was detectable, the peak height over noise was $6.2 \pm 1.9$ for the near-zero primary velocity case, while it was $2.6 \pm 1.0$ for the even primary velocity case, $2.3 \pm 1.0$ for the most red-shifted primary velocity case, $1.9 \pm 0.7$ for the most blue-shifted primary velocity case, and $2.7 \pm 1.0$ for the random primary velocity case. The simulation results show a significant amount of scatter around these averages, even at a single value of $\gamma$, as seen in Figure~\ref{guassianfitstorandomMfigure}. We found no significant relationship between Gaussian height over noise and the mean orbital phase, standard deviation of the orbital phases, or the standard deviation of the $|\sin(2\pi M_i)|$ values, though. We also ran 100 simulations with the same orbital phases ($\gamma = 0.67$) and primary velocities (near-zero) to see how much of the scatter could be explained by white noise. The Gaussian heights from these simulations had a standard deviation of only 0.3, less than the scatter in any of the five primary velocity groups. This implies that with S/N of 2860 per pixel per epoch, the structured noise related to the combination of orbital phases dominates over random Gaussian noise. This is consistent with findings from~\citet{Buzard2020} and \citet{Finnerty}.

The lower panels of Figure~\ref{guassianfitstorandomMfigure} show the widths of the Gaussian fits. The gray dashed line in each subplot shows the average velocity precision of the upgraded NIRSPEC data on which these simulations were based, at 3.1 km/s. The widths of the near-zero primary velocity simulations show an interesting trend. At large values of $\gamma$, the widths are quite small and do not have much variation. The widths increase toward intermediate $\gamma$ values in both magnitude and degree of variation. The representative near-conjunction simulation, shown by the blue star near $\gamma = 0$, has by far the largest width. The trend in width magnitude reflects the fact the conjunction epochs have very little constraining power on $K_p$ while quadrature epochs are the most effective for constraining $K_p$. The degree of variation in Gaussian width at intermediate values of $\gamma$ as opposed to large or small values can also be explained. While there is only one combination of epochs each that will give a $\gamma$ value of 0 (all at conjunction) or 5 (all at quadrature), there are many different combinations of epochs that could result in an intermediate value of $\gamma$. For instance, the $\gamma$ values of 1000 sets of 5 epochs with orbital phases pulled from a uniform distribution from 0 to 1 form an approximately Gaussian shape with a mean of 0.64 and a standard deviation of 0.14. With more cases at intermediate values of $\gamma$, there will be more variation as some of them will provide better constraints on $K_p$ than others.   % why is gamma centered on 3.2 rather than 2.5?

The four primary velocity groups other than the near-zero group do not show the same strong relationship between the Gaussian widths and $\gamma$. The most blue-shifted and most red-shifted primary velocity groups do show some evidence of a corner where the widths increase below a certain $\gamma$ value but this behavior is not nearly as strong as the trend in the near-zero primary velocity case. We suspect that the higher planetary peaks in the near-zero group are better fit by a Gaussian, meaning that the widths are more representative of the true planetary detection peak than for the other primary velocity groups. This can be corroborated by the mean $R^2$ value of the detected planetary peaks in each $v_{pri}$ group, which compares the goodness of the Gaussian fit to that of a horizontal line at the mean of the simulated log likelihood curve. The mean $R^2$ values of the near-zero, most blue-shifted, even, most red-shifted, and random primary velocity groups are 0.69, 0.25, 0.34, 0.24, and 0.36, respectively. The especially low $R^2$ values of the four primary velocity groups other than the near-zero one reflect the high levels of structured noise. They also support our conjecture that the widths of those four groups show less dependence on $\gamma$ than the near-zero primary velocity group because the Gaussian fits are not accounting for the planetary peak structure as accurately.

Since neither the most blue-shifted (most negative) nor the most red-shifted (most positive) primary velocity groups were able to strongly detect the planetary signal, going forward, we will refer to both as the ``largest absolute primary velocity group." Doing so allows us to focus on the magnitude of velocity separation between the stellar signal and the telluric frame, rather than the direction in which the stellar signal has moved. 

Collectively, these results show that a set of 5 epochs with randomly selected orbital phases will have the best chance of showing a strong detection of the planet if they are taken during times when the system's velocity is canceled out by the Earth's velocity in the direction of the system. The closer these epochs are to quadrature, the better the data will be able to constrain the value of $K_p$. Further, we would expect that obtaining data from both quadrature positions ($M = 0.25$ and $0.75$) would be better for constraining $K_p$ than data at just one quadrature position because having data at both quadrature positions would give us access to different velocity shifts relative to the telluric frame and so collectively more complete wavelength coverage of the planetary spectrum. \citet{Finnerty} showed that a larger spectral grasp can drastically increase detection significance; doubling the grasp increased the significance by nearly a factor of 2. These predictions should be useful for the planning of future multi-epoch observations.

\section{Magnitude of $K_p$}  \label{Section:KpMagnitude}

All simulations presented in Section~\ref{Section:simulations} considered data sets generated with a $K_p$ of 75 km/s. We showed that near-zero primary velocity epochs allow for the strongest planetary detections. However, we might expect to encounter a challenge, especially as $K_p$ decreases, with setting the primary velocity to 0, or in other words, allowing very little velocity shifting between the stellar and telluric spectra. If the stellar spectrum is not velocity shifted relative to the telluric spectrum, as $K_p$ decreases, the planetary lines will not be able to stray much from the telluric spectrum either. 

The value of $K_p$ is set by both the semi-major axis and the orbital inclination. A decrease in $K_p$ due to a larger semi-major axis would be accompanied by a colder planetary effective temperature, while a decrease in $K_p$ solely due to a smaller inclination would not affect the planetary temperature. While either decrease in $K_p$ would limit the separation between planetary and telluric lines in near-zero primary velocity epochs, cooler planetary atmospheres could further complicate the issue. While hot Jupiter $L$ band spectra are dominated by water features, their water is much hotter ($\gtrsim1000$ K) than water in the telluric spectrum ($\sim300$ K), resulting in very different spectral line shapes, position, and relative contrast. However, as the planetary effective temperature decreases, its spectrum will more and more resemble that of the Earth's. Then, with neither very different temperatures altering the shape of the planetary spectrum from the telluric spectral shape or much velocity shifting off of the telluric spectrum, primary velocity near-zero epochs may no longer be as useful. Such cool planets will require large epoch number data sets to be detected.    

To test how varying $K_p$ affects our simulation results, we generate 100 simulations with near-zero primary velocities and randomly selected orbital phases with $K_p$ values of 37.5, 75, and 150 km/s. We maintain a common planetary effective temperature in order to examine the effects on hot Jupiter detectability with a changing orbital inclination, not on the detectability of planets with different effective temperatures due to different semi-major axes. In analyzing these simulated data sets, we calculated the planetary log likelihoods vs. $K_p$ from $-250 \leq K_p \leq 250$ km/s to allow for sufficient parameter space to robustly constrain the 150 km/s detections.

\begin{figure}
    \centering
    \noindent\includegraphics[width=21pc]{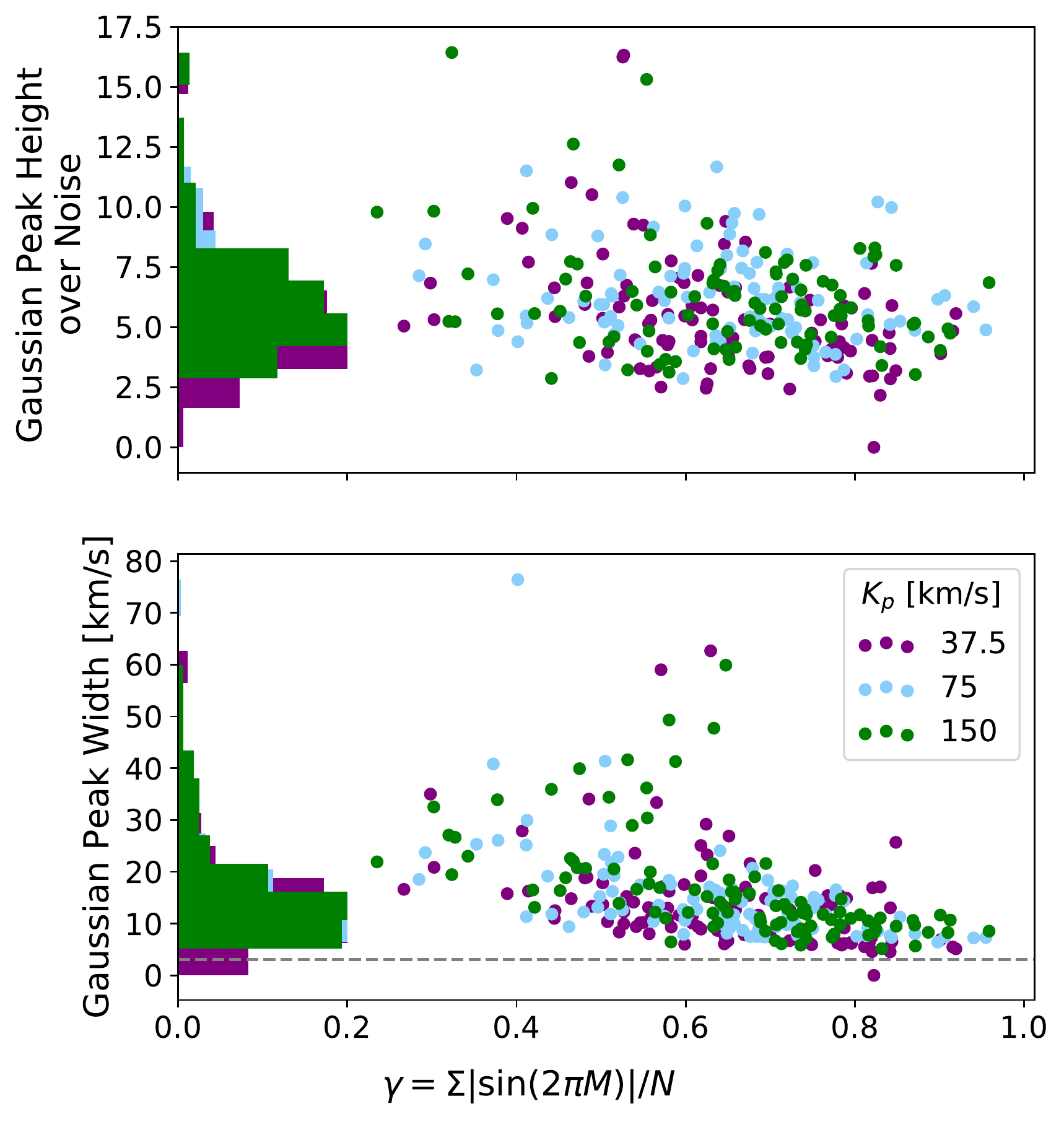}
    \caption{ Results of Gaussian fits to simulations with near-zero primary velocities, a random selection of 5 orbital phases, and different values of $K_p$. Points in purple represent simulations with a $K_p$ of 37.5 km/s, light blue points have a $K_p$ of 75 km/s, and green points have a $K_p$ of 150 km/s. Of the 300 simulations, only one was unable to detect the planetary signal; it was one of the $K_p = 37.5$ km/s simulations.   }
    \label{Figure:gaussiandifferentKps}
\end{figure}

Figure~\ref{Figure:gaussiandifferentKps} shows the Gaussian heights relative to the noise and the Gaussian widths of these simulations. Of all 300 simulations, only one of the $K_p = 37.5$ km/s cases was unable to fit the peak within $1\sigma$. 

We used a Kolmogorov–Smirnov (KS) test to determine whether there was any statistical difference between the Gaussian heights over noise and widths of the sets of simulations with different values of $K_p$. Two-tailed p-values between the $\gamma$ values of the 37.5, 75, and 150 km/s simulations were 0.68 (37.5 vs. 75 km/s), 0.34 (37.5 vs. 150 km/s), and 0.34 (75 vs. 150 km/s). None of these p-values are small enough to justify rejecting the null hypothesis that the 100 $\gamma$ values of each of the $K_p$ cases were pulled from the same distribution. We know the null hypothesis to be true in this case; all 300 $\gamma$ values were pulled from the same distribution, the conversion of the 5 $M$ values uniformly pulled from 0 to 1 through Equation~\ref{Equation:gamma}. This result then helps to validate the use of the KS test. 

The KS p-values between the Gaussian heights over noise were 0.0030 (37.5 vs. 75 km/s), 0.031 (37.5 vs. 150 km/s), and 0.89 (75 vs. 150 km/s). The Gaussian width p-values were 0.031 (37.5 vs. 75 km/s), 0.069 (37.5 vs. 150 km/s), and 0.56 (75 vs. 150 km/s). The peak heights and widths of the 75 and 150 km/s cases can both be assumed to be pulled from the same parent distributions. On the other hand, KS tests reject the hypotheses that the 37.5 km/s heights are pulled from the same parent distribution as the 75 km/s heights at the $3.0\sigma$ level and from the same parent distribution as the 150 km/s heights at the $2.2\sigma$ level. They reject a common parent distribution between the 37.5 and 75 km/s Gaussian widths at the $2.1\sigma$ level and between the 37.5 and 150 km/s widths at the $1.8\sigma$ level. While these levels of statistical rejection of the null hypotheses are mostly in the ``weak" to ``moderate" support categories \citep[e.g., ][]{Gordon2007}, when considered alongside the means and standard deviations of the Gaussian heights of each distribution, they do start to show weaker planetary detectability at lower values of $K_p$. The 37.5, 75, and 150 km/s sets of simulations have average Gaussian peak heights over the noise level of $5.5 \pm 2.4$, $6.2 \pm 1.9$, and $6.2 \pm 2.3$, respectively. While the 37.5 km/s peak heights were moderately lower than the 75 km/s and 150 km/s peak heights here, they are still on average significantly higher than the 75 km/s peak heights measured from the primary velocity groups not near zero, as described in Section~\ref{Section:simulations}.

For hot planets, we find that near-zero primary velocity epochs allow for stronger detections than other combinations of primary velocities even as $K_p$ gets quite small. This result may be challenged as we look to cooler planetary atmospheres which will have a higher degree of spectral similarity to our own telluric spectrum.

\section{Comparison to Data} \label{Section:ComparetoDataSection}

We were interested to test whether our prediction that epochs with primary velocities near zero would give stronger detections than other samples of primary velocities would hold up against previous NIRSPEC observations. NIRSPEC has been used to obtain multi-epoch detections of exoplanets dating back to 2011; the first set of which were published by \citet{lockwood} (Tau Bo\"{o}tis b). Unfortunately, there are not enough NIRSPEC epochs for any one system to be able to test the effectiveness of different primary velocity groupings. Therefore, in order to test our predictions, we combine epochs from different targets. From our archive of NIRSPEC observations, we compile the five epochs with the primary velocities nearest zero and the five epochs with the largest absolute primary velocities (which happen to all be in the ``most blue-shifted"/most negative category). These epochs are from Tau Boo b, HD187123b, 51 Peg b, and KELT2Ab. All of these planets are on orbits that can be approximated as circular. 

In order to combine all epochs we need to perform a change-of-base so that the epochs reflect a single Keplerian line-of-sight orbital velocity $K_p'$. We denote all true parameters from the different systems without a prime, and all parameters of the fictitious combined system with a prime ($'$). The primary ($v_{pri}$) and secondary ($v_{sec}$) velocities are encoded in the data and so cannot be altered. For a single system, $v_{pri}$ is variable because of the changing barycentric velocity in the direction of the system, but here, the variability in $v_{pri}$ can account for both changing barycentric velocities, and the different systemic velocities of the different targets. $K_p'$ must be large enough to account for all the values of $v_{sec} - v_{pri}$; we set it to 150 km/s. Then, rearranging Equation~\ref{Equation:secondaryvelocity} for the secondary velocity, we get

\begin{equation}
    M' = \frac{1}{2\pi}\arcsin{\frac{v_{sec} - v_{pri}}{ K_p'}}.
\end{equation}

\begin{deluxetable*}{lcccccccc} 
\tablewidth{0pt}
\def\arraystretch{1}
\tablecaption{Target Information}
\tablehead{Target & $K_p$ & $K_p$ Ref. & Period  & Period Ref.  & $T_0$ & $T_0$ Ref. & $v_{sys}$ & $v_{sys}$ Ref. \\
 & [km/s] &  & [days]  &   & [JD] &  & [km/s] }
\startdata
HD187123 &  53 $\pm$ 13 & (1)  &  3.0965885$^{+0.0000051}_{-0.0000052}$ &(1) &     2454343.6765$^{+0.0064}_{-0.0074}$ & (1) & -17.046 $\pm$ 0.0040 & (7) \\
Tau Boo     & 111 $\pm$ 5& (2)  & 3.312433 $\pm$ 0.000019  & (3) & 2455652.108 $\pm$ 0.004 & (3) &  -16.03 $\pm$ 0.15 & (9) \\
KELT2A    &  148 $\pm$ 7 & (4)  & 4.1137913 $\pm$ 0.00001 & (5)    & 2455974.60338$^{+0.00080}_{-0.00083}$ & (5) & -47.4 $\pm$ 0.6 & (8) \\
51 Peg   & 133$^{+4.3}_{-3.5}$ & (6)   & 4.2307869$^{+0.0000045}_{-0.0000046}$  & (6)    & 2456326.9314 $\pm$ 0.0010 & (6) & -33.165 $\pm$ 0.0006 & (7)
%% these aren't currently the P, To i have in the idl run files
\enddata
\label{targetstable}
\tablecomments{We assume circular orbits for all of these targets.}
\tablerefs{(1) \citealt{Buzard2020}, (2) \citealt{lockwood}, (3) \citealt{Brogi2012}, (4) \citealt{Piskorz2018}, (5) \citealt{Beatty2012}, (6) \citealt{Birkby2017}, (7) \citealt{Gaia2018}, (8) \citealt{Gontcharov2006}, (9) \citealt{Nidever2002} } 
\end{deluxetable*}

\begin{deluxetable*}{lccccccc} 
\tablewidth{0pt}
\def\arraystretch{1}
\tablecaption{Epoch Information}
\tablehead{Target & Obs. Date & $t_{\mathrm{obs}}$ [JD] & $v_{bary}$ [km/s] & $v_{pri}$ [km/s] & $M$ & $v_{sec}$ [km/s]  & $M'$\tablenotemark{a} }
\startdata
\sidehead{\textbf{Primary Velocity Near Zero}}
Tau Boo    & 2011, May 21 & 2455702.85  &  -17.34 &  1.31 & 0.319 &102.08 & 0.117 \\
HD187123    & 2013, Oct 27  & 2456592.76 & -17.44  &0.40 & 0.310 & 49.70 & 0.053  \\
HD187123    & 2013, Oct 29 & 2456594.74 &  -17.50 & 0.45 & 0.949 &  -16.27 & 0.982  \\
51 Peg & 2013, Nov 07    & 2456603.86  &    -21.27       & -11.90 &0.455 &25.46 & 0.040 \\
HD187123    & 2017, Sep 07 & 2458003.77 & -10.15  & -6.90 & 0.977 & -14.45 & 0.992  \\
\sidehead{\textbf{Largest Absolute Primary Velocities}}
51 Peg   & 2011, Aug 10 & 2455783.96  & 15.77 & -48.94 & 0.661& -161.73 &0.865 \\
51 Peg & 2014, Sep 04    &  2456905.04 & 5.43  & -38.59 & 0.643 & -142.49 &0.878  \\
KELT2A    & 2015, Dec 01 &  2457357.89 &  11.77 & -59.18 &0.256  & 88.70 &0.223  \\
KELT2A    & 2015, Dec 31 & 2457387.97 & -3.62  & -43.77 & 0.567 & -104.39 & 0.934  \\
KELT2A    & 2016, Dec 15  &2457738.10 &  4.79 & -51.68 & 0.680  &-185.65 &0.824 
\enddata
\label{epochtable}
\tablenotetext{a}{$M'$ is the orbital phase, $M$, reflecting the change-of-base to $K_p' = 150$ km/s so that the epochs can be combined. }
\end{deluxetable*}

Table~\ref{targetstable} gives the true parameters ($K_p$, $P$, $T_o$, $v_{sys}$) from the four target systems. Table~\ref{epochtable} gives the information about the specific dates we are considering. Because these systems have different expected stellar and planetary spectra, we use different spectral templates to cross correlate each epoch of data, and then use the new change-of-base $M'$ values to combine the log likelihood curves generated from the two dimensional cross correlations. The data reduction and stellar and planetary spectral template used for cross correlation for each of the sources is described in the Appendix. The five log likelihood curves that make up the two primary velocity groups are shown in Figure~\ref{mlcombinedatafigure}.  

\begin{figure}
    \centering
    \noindent\includegraphics[width=21pc]{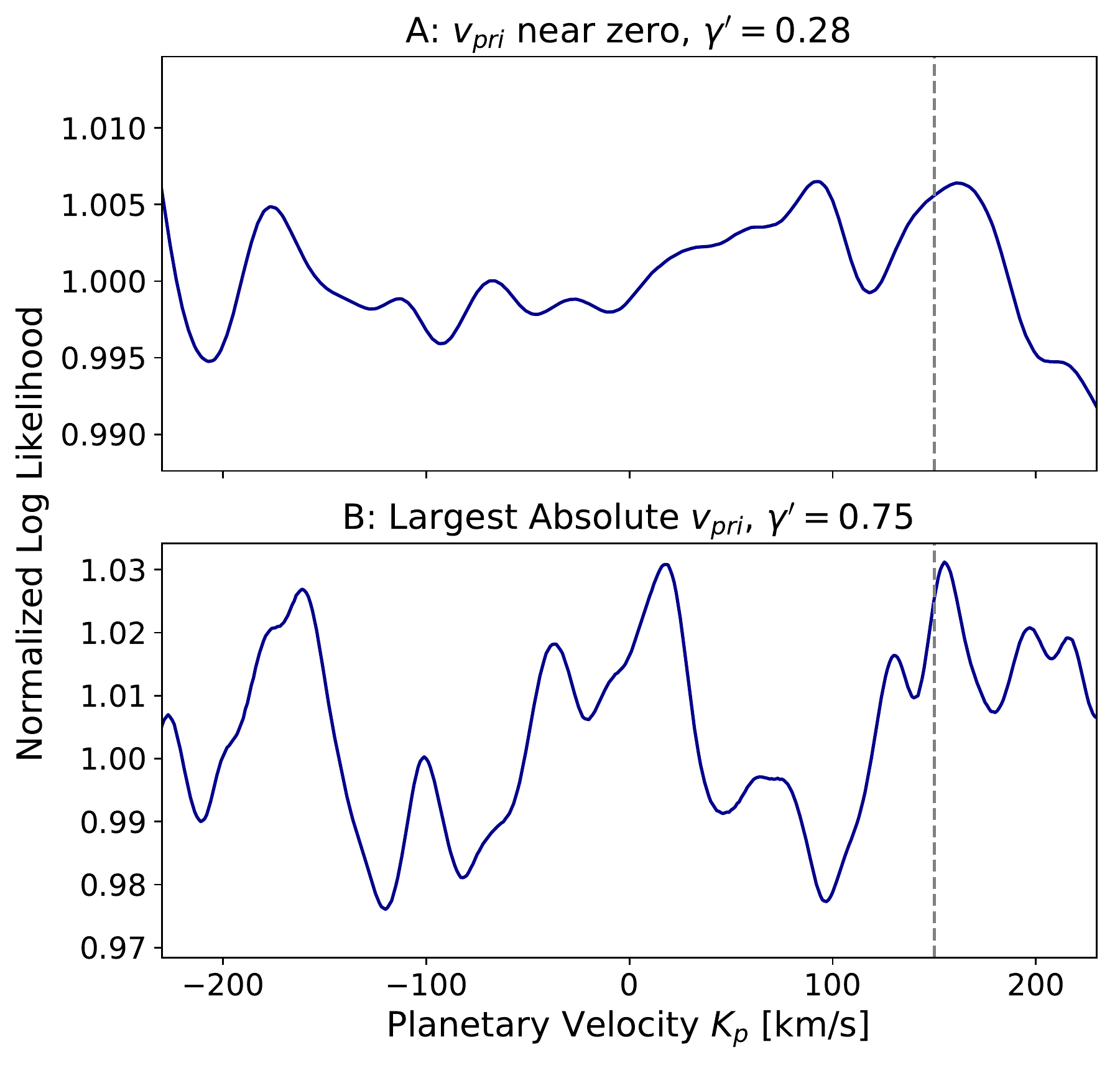}
    \caption{ Normalized log likelihood vs. $K_p$ from pre-upgrade NIRSPEC data epochs. The top panel shows the combination of 5 epochs with primary velocities nearest zero and the bottom panel shows the combined epochs with the largest absolute primary velocities.     }
    \label{DataFigure}
\end{figure}

Aside from the different planetary and stellar spectral models used for each epoch, there are a few other differences between these combinations of data epochs and the predictions for the near-zero and largest absolute primary velocity groups in Section~\ref{Section:simulations}. First, all 10 of the data epochs were taken prior to the NIRSPEC upgrade, while the simulations considered post-upgrade NIRSPEC specifications. These differences affect the total S/N, the instrument resolution, wavelength coverage, and wavelength range covered (see Appendix). Second, the primary velocity groups are not defined as strictly here as they were in Section~\ref{Section:simulations}. This is simply due to the availability of data epochs. While the near-zero primary velocity group in Section~\ref{Section:simulations} chose primary velocities from -2 to 2 km/s, the data near-zero primary velocity group have primary velocities ranging from -11.9 to 1.3 km/s. The simulated largest absolute (most blue-shifted/most negative) primary velocity group pulls $v_{\mathrm{pri}}$ values from -30 to -28 km/s, while the data largest absolute primary velocity group ranged from -59.2 to -38.6 km/s. 

In Section~\ref{Section:KpMagnitude}, we found no statistical difference between the Gaussian heights relative to the noise or Gaussian widths of the sets of simulations with $K_p$ values of 75 vs. 150 km/s. Therefore, the fact that these data are set up with a $K_p$ of 150 km/s should not be one of the factors differentiating these results from the $K_p = 75$ km/s simulations.

Figure~\ref{DataFigure} shows the normalized log likelihoods of the five epochs with primary velocities near zero and the five epochs with the largest absolute primary velocities. When fit with Gaussians in the same way as the analysis shown in Figures~\ref{guassianfitstorandomMfigure} and \ref{Figure:gaussiandifferentKps}, the primary velocity near-zero case can be fit by a Gaussian at $157 \pm 15$ km/s with a height over the noise of 1.7, while the largest absolute primary velocity case fit gives a value of $180 \pm 34$ km/s with a height over the noise of 1.4.

We calculate $\gamma '$ values for the data combinations of epochs as 0.28 for the near-zero primary velocity group and 0.75 for the largest absolute primary velocity group. While the near-zero primary velocity case has a lower $\gamma '$ value, its fit is slightly more accurate and higher relative to the noise than the largest absolute primary velocity case.

\begin{figure}
    \centering
    \noindent\includegraphics[width=21pc]{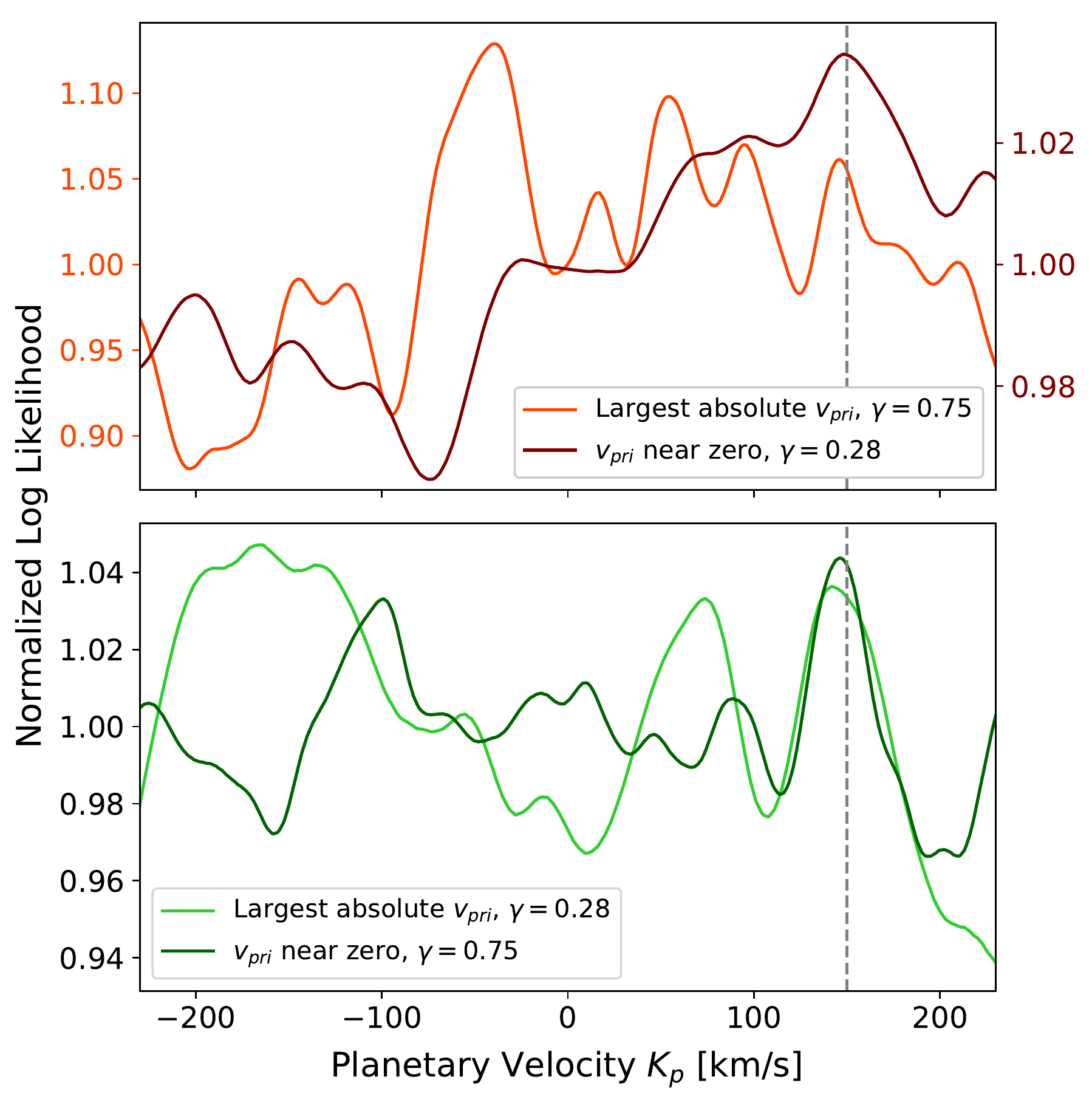}
    \caption{ Results of simulations with the primary velocities and orbital phases of the data epochs combined in Figure~\ref{DataFigure}. The top panel has the correct grouping of primary velocities and orbital phases and the bottom panel swaps the primary velocities and orbital phases to test whether the difference in detection strengths can be said to be mostly from the different primary velocity groups or whether the combination of orbital phases also had a large effect on the detection strengths. Unlike Figures~\ref{simulationsfigure}-\ref{Figure:gaussiandifferentKps} and~\ref{Figure:Nepochs}, these simulations are for pre-upgrade NIRSPEC data. } %nirspec1, ips from HD187123 5 nirspec 1 epochs, 1 jupiter radius, 1 solar radius }
    \label{simulatingdatafigure}
\end{figure}

We ran simulations with the exact primary velocities and $M$s (and $\gamma$s) from the data near-zero and largest absolute primary velocity groups to determine (1) if the primary velocity groups still showed a similar trend when they were not defined as strictly as in the simulations in Section~\ref{Section:simulations}, (2) if the near-zero primary velocity epochs were more effective with pre-upgrade NIRSPEC settings as well as with post-upgrade NIRSPEC settings, and (3) if this trend in the data is based on the different combinations of epoch orbital phases or can be assigned to primary velocity differences. For these simulations, we only consider a single planetary and stellar spectral model, defined in Section~\ref{Section:spectralmodels}, for each epoch. Thus, we do not expect them to appropriately reproduce the off-peak structure in Figure~\ref{DataFigure}, but they should allow us to answer the questions listed above. 

The results of the simulations are shown in Figure~\ref{simulatingdatafigure}, with the primary velocities and orbital phases of the data in the top panel. We show these simulated log likelihood curves with separate y-axes because the structured noise in the largest absolute primary velocity simulation is at a much higher level than that in the near-zero primary velocity case. Gaussian curves find fits for the near-zero $v_{pri}$ and $M$s (in maroon) of $138 \pm 81$ km/s with a height relative to the noise of 4.4 and the largest absolute $v_{pri}$ and $M$s (in orange) of $36 \pm 79$ km/s with a height of 1.1, which would not be considered a detection as it is more than $1\sigma$ away from 150 km/s. These simulations then do indeed agree that the data near-zero primary velocity epochs have a better chance of detecting the planet. This validated that the near-zero $v_{pri}$ epochs are more effective with both pre- and post-upgrade NIRSPEC and that they are still preferable to larger absolute value primary velocity epochs even if not as strictly defined to $-2 \leq v_{pri} \leq 2$ km/s. 

We do note that the simulations predict a much larger improvement going from the largest absolute primary velocity case to the near-zero primary velocity case (1.1 to 4.4) than was seen in the data (1.4 to 1.7). Recall that the simulated data sets are generated assuming that all non-saturated tellurics can be perfectly corrected. This is likely not the case in the real data. Because the planetary velocities are closer to the telluric frame in the near-zero primary velocity group (both because of the near-zero primary velocities themselves and because of the smaller $\gamma$, see Table~\ref{epochtable}), the planetary spectral lines are closer to the corresponding telluric lines than in the largest absolute primary velocity group. The $L$ band wavelengths covered are dominated by water features at hot Jupiter temperatures, and while this water is much hotter than telluric water, the closer overlap between its features and the imperfectly corrected telluric water features could be responsible for hampering the near-zero primary velocity case detection significance in the real data.      

\begin{deluxetable}{cccccc} 
\tablewidth{0pt}
\def\arraystretch{1}
\tablecaption{Gaussian Fits to Data}
\tablehead{Type & $v_{pri}$ Group & $\gamma$  & $\mu$ & $\sigma$ & A }
\startdata
Data   & Near Zero & 0.28  & 157  & 15 & 1.7 \\
Data   & Largest Absolute  & 0.75  & 180  & 34 & 1.4 \\
Simulation   & Near Zero &  0.28 & 138  & 81 & 4.4 \\
Simulation   & Near Zero  & 0.75  & 146  & 9 & 3.3 \\
Simulation   & Largest Absolute & 0.28  & 145  & 10  & 1.0 \\
Simulation\tablenotemark{a}   & Largest Absolute & 0.75  & 36  & 79 & 1.1
\enddata
\label{Table:datasimresults}
\tablenotetext{a}{This would be considered a non-detection because the set $K_p'$ of 150 km/s is more than $1\sigma$ from the Gaussian mean. }
\end{deluxetable}

Using these simulations, we next test whether the improvement of the primary velocity near-zero epochs over the largest absolute primary velocity epochs was in fact due to the primary velocity differences, or if it was made by the different combinations of orbital phases. To do so, we ran simulations with the primary velocities and orbital phases of the data swapped. The results of these swapped simulations are in the lower panel of Figure~\ref{simulatingdatafigure}. The simulation with the near-zero primary velocities but orbital phases of the largest absolute primary velocity epochs (in dark green) can be fit as $146 \pm 9$ km/s, with a relative height of 3.3 and the simulations with the largest absolute primary velocities but the near-zero orbital phases (in light green) was fit as $145 \pm 10$ km/s and a height over noise of 1.0. These simulations would both qualify as detections, but the one with near-zero primary velocities is stronger. Table~\ref{Table:datasimresults} lists the Gaussian parameters of the two data and four simulated log likelihood curves. 

%\textcolor{blue}{common priamry velocities with different gammas.  }

Figure~\ref{guassianfitstorandomMfigure} saw no real trend in the Gaussian heights relative to the noise as a function of $\gamma$ in any of the primary velocity groups. This is seen in the data simulations as well. The near-zero primary velocity data epochs had orbital phases corresponding to a $\gamma$ of 0.28, while the largest absolute primary velocity data epochs had orbital phases corresponding to a $\gamma$ of 0.75. The data near-zero primary velocities gave a Gaussian height of 4.4 when combined with the $\gamma = 0.28$ orbital phases vs. 3.3 with the $\gamma = 0.75$ epochs. The largest absolute data primary velocities showed a height relative to the noise of 1.0 with the $\gamma = 0.28$ epochs and was not detected with the $\gamma = 0.75$ epochs. These results support our finding that the primary velocities of the data epochs had a stronger effect on the detection strength than the positions of the orbital phases. It could be that the velocity separation given by epochs far from conjunction will be important though, especially for near-zero primary velocity epochs, when residual telluric features cannot be perfectly corrected from the data. 
%near zero epochs were responsible for the stronger detection, rather than the specific combinations of orbital phases accompanying the near-zero primary velocities in the data epochs.   

Figure~\ref{guassianfitstorandomMfigure} also showed a fairly significant correlation between increasing $\gamma$ and decreasing Gaussian width in the near-zero primary velocity group. This too is found in these new data simulations: the near-zero primary velocities found a Gaussian width of 81 km/s for a $\gamma = 0.28$ data set and 9 km/s for $\gamma = 0.75$.  

\section{Number of Epochs} \label{Section:EpochNumbers}

\begin{figure}
    \centering
    \noindent\includegraphics[width=21pc]{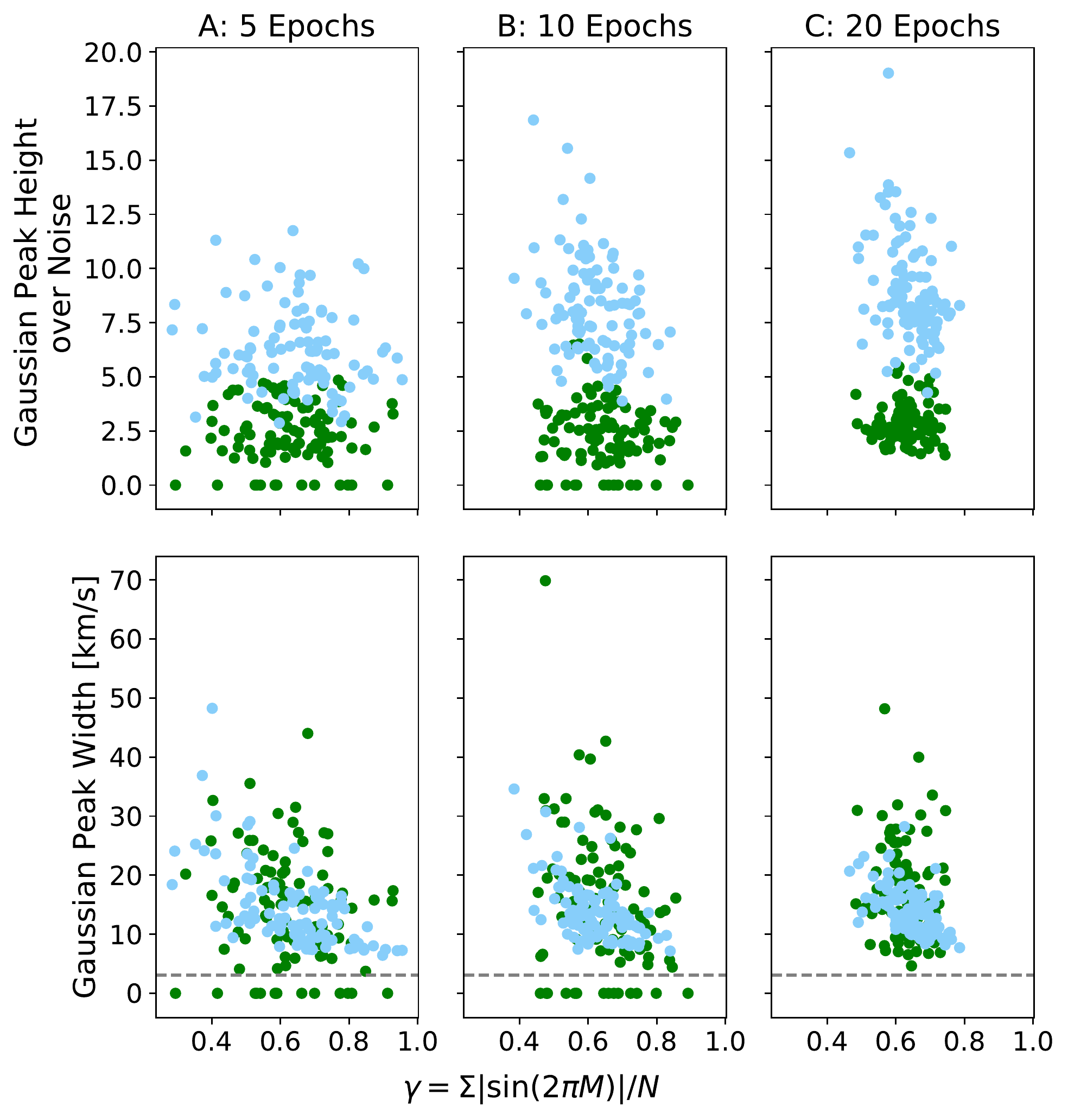}
    \caption{ Results of Gaussian fits to 5, 10, and 20 epoch simulations with random and near-zero primary velocities and random orbital phases. The near-zero primary velocity simulations are shown in light blue and the random primary velocity simulations are shown in green. Of the 600 simulations, only 14 of the 5-epoch and 14 of the 10-epoch random primary velocity simulations were unable to detect the planetary signal.     } %nirspec1, ips from HD187123 5 nirspec 1 epochs, 1 jupiter radius, 1 solar radius }
    \label{Figure:Nepochs}
\end{figure}

The simulations thus far have all considered 5 epochs of data. We were interested to see what increasing the number of simulations would do to the detection strengths of the different primary velocity groups. To do this, we compared the random and near-zero primary velocity groups with 5, 10, and 20 epochs. We maintain a constant total S/N per pixel in a simulation across all of the epochs. The S/N per epoch then decreases with increasing epoch number from 2860 (5 epochs), to 2020 (10 epochs), to 1430 (20 epochs). Figure~\ref{Figure:Nepochs} shows the results of these epoch number simulations, with near-zero primary velocity epoch simulations shown in light blue and random primary velocity epoch simulations shown in green. These Gaussian results are plotted with respect to $\gamma$.

Of the six groups (near-zero and random primary velocity groups with 5, 10, and 20 epochs), all simulations were able to detect the planetary signal except 14 of each the 5- and 10-epoch random primary velocity simulations. The mean Gaussian heights over noise of the detected planetary signals for the near-zero primary velocity groups are $6.2 \pm 1.9$, $8.2 \pm 2.3$, and $8.9 \pm 2.3$, for the 5, 10, and 20 epoch simulations, respectively. For the random primary velocity epochs, the means are $2.7 \pm 1.1$ for the 5 epoch case, $2.6 \pm 1.1$ for the 10 epoch case, and $2.8 \pm 0.8$ for the 20 epoch case. Interestingly, we see that the two populations actually appear to diverge as the number of epochs increases, rather than converge as we had expected. The random primary velocity simulations maintain a similar Gaussian height over noise as the number of epochs increases. KS statistics tell us that the 5 and 10 epoch random primary velocity results can be assumed to have been pulled from the same distribution with very high confidence ($p = 0.89$). The 20 epoch random $v_{pri}$ group can be assumed to be pulled from a different distribution from each of the 5 and 10 epoch random $v_{pri}$ groups at a $3.7\sigma$ confidence level. We can see in Figure~\ref{Figure:Nepochs}, and in the reported mean, that the 20 epoch simulations have much less variance in peak height over noise than the 5 and 10 epoch random $v_{pri}$ populations. Visually, and through the reported means, unlike the random primary velocity cases, the near-zero primary velocity heights over the noise increase from 5 to 10 epochs, and then seem to stabilize from 10 to 20 epochs. While the 5 and 10 epoch near-zero simulation heights can be said to be from different parent distributions at a $5.4\sigma$ confidence level, the 10 and 20 epoch near-zero distributions are only distinct at a $2.3\sigma$ confidence level. As we will address later, these values do not account for the fact that in addition to being generated from simulations with different numbers of epochs, these heights are apparently pulled from different $\gamma$ distributions. If this were accounted for, we would expect even more commonality between the near-zero heights from 10 and 20 epoch simulations.  

If we compare the Gaussian heights over the noise level from the random and near-zero primary velocity groups at 5, 10, and 20 epochs, respectively, we find that two-tailed p-values measured from the Kolmogorov-Smirnov statistic decrease from $5.1 \times 10^{-30}$ with 5 epochs to $2.8 \times 10^{-40}$ with 10 epochs to $9.5 \times 10^{-44}$ with 20 epochs. While the heights of planetary detections resulting from the random and near-zero primary velocity groups can always be said to be pulled from statistically distinct populations, the level at which this claim can be made increases by orders of magnitude as the number of epochs in the simulation increases. The larger jump from the 5 to 10 epoch p-values versus the 10 to 20 epoch p-values reflects the leveling off of the peak heights over noise from 10 to 20 epochs.

As in Figure~\ref{guassianfitstorandomMfigure}, the near-zero primary velocity groups widths, at every number of epochs, show a decreasing trend in both magnitude and variability with increasing $\gamma$. The random primary velocity group widths do not show this trend as strongly, likely indicative of the Gaussian models not fitting the planetary peak structure as well as in the near-zero primary velocity case.

Another thing we see in Figure~\ref{Figure:Nepochs} is that as the number of epochs increases, the distribution of $\gamma$ values narrows. In fact, while the mean $\gamma$ distribution across 1000 5-epoch simulations would be $0.64 \pm 0.14$, for 1000 10- and 20-epoch simulations, it would be $0.64 \pm 0.10$ and $0.64 \pm 0.07$, respectively. We might then expect the distribution of random primary velocities to be narrowing with increasing epoch number too. To see this directly, we can define a parameter $\beta$ that quantifies the combination of primary velocities in a simulation, similarly to how $\gamma$ quantifies the combination of orbital phases.
\begin{equation}
    \beta = \frac{1}{30N}\Sigma |v_{pri}|
\end{equation}
We divide by 30 to normalize by the maximum absolute primary velocity of our simulation. Then, the mean of 1000 $\beta$ values for 5, 10, and 20 epoch simulations would be $0.63 \pm 0.14$, $0.64 \pm 0.09$, and $0.64 \pm 0.07$ as well. This could explain why the random primary velocity simulations do not benefit from more epochs to the same extent that the near-zero primary velocity epochs do. As the number of epochs increases, the $\beta$ distribution from which the random primary velocities are drawn is increasingly pulled away from the optimal near-zero case. While the heights of the Gaussian peaks from the near-zero primary velocity group increase from 5 to 10 epoch simulations, the 10 epoch random primary velocity simulations would have a slightly worse placement of primary velocities than the 5 epoch case. The larger number of epochs and worse placement balance each other out so that the Gaussian heights remain comparable at different epoch numbers.  

From these analyses, we have seen that the average planetary detection from near-zero primary velocity epochs grew by a factor of 1.3 from 5 to 10 epoch simulations and nearly leveled out from 10 to 20 epoch simulations. The planetary detectability from random primary velocity epochs was nearly comparable at 5, 10, and 20 epochs, though there were no non-detections with 20 epoch simulations and a 14\% non-detection rate at 5 and 10 epochs. As we saw above, the near-zero primary velocity epochs can be better modeled by a Gaussian than the random primary velocity epochs and as a result, their Gaussian widths show a stronger dependence on $\gamma$.

\section{Stellar Properties} \label{Section:stellarproperties}

\begin{figure*}
    \centering
    \noindent\includegraphics[width=42pc]{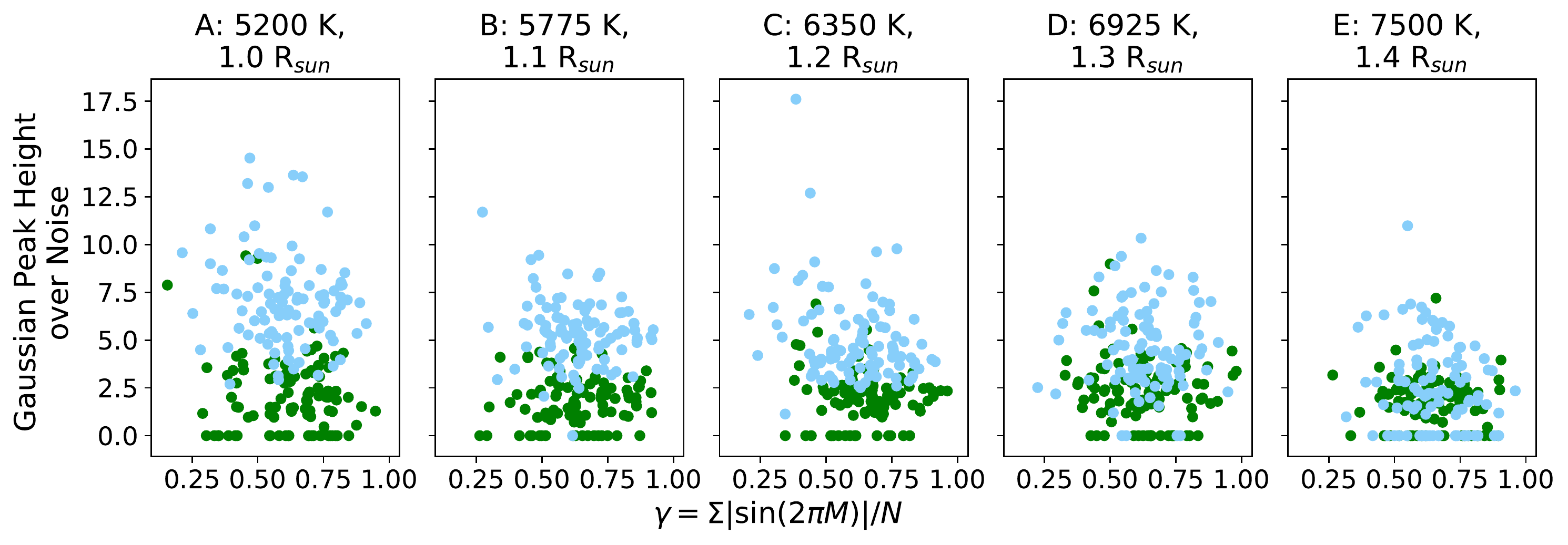}
    \caption{   Results of 5 epoch simulations with different host stellar temperatures and radii. Near-zero primary velocity simulations are shown in light blue and random primary velocity simulations are shown in green. The numerical results are reported in Table~\ref{Table:stellartemps}.   }
    \label{Figure:stellartemps}
\end{figure*}

We next investigated how the properties of the host star affect the optimal primary velocity observing strategies. Could it be that the near-zero primary velocity observing strategy works well with the 5815 K stellar model assumed because this star has strong lines corresponding to strong telluric features that are removed by telluric masking when these spectra are aligned? If so, will the near-zero primary velocity observing strategy be as effective for other stellar temperatures? To test the generalizability of the near-zero primary velocity approach, we ran 100 near-zero and 100 random primary velocity simulations with five stellar models ranging from 5200 to 7500 K, to cover the F and G spectral types. We increased the stellar radius along with the temperature, but maintained a constant metallicity and surface gravity. We also used the same planetary model in each case. Results from these simulations are outlined in Table~\ref{Table:stellartemps} and shown in Figure~\ref{Figure:stellartemps}.    

\begin{deluxetable*}{cccccccc} 
\tablewidth{0pt}
\def\arraystretch{1}
\tablecaption{Stellar Temperature Simulation Results}
\tablehead{Temperature [K] & Radius [$R_{\sun}$] & \% detected  & Peak Height/Noise & \% detected  & Peak Height/Noise & $p$-value\tablenotemark{a} & $n\sigma$\tablenotemark{a} \\ & & Near-zero & Near-zero  & Random & Random & & }
\startdata
5200 & 1.0  &  100 & $7.0 \pm 2.4$  & 75  & $2.8 \pm 1.7$ & $9.3 \times 10^{-31}$ & \\
5775  & 1.1 & 99  & $5.4 \pm 1.6$  & 82  & $2.3 \pm 1.0$ & $5.1 \times 10^{-30}$ &  \\
6350  & 1.2 & 100  & $5.0 \pm 2.3$  & 84  & $2.4 \pm 1.0$  & $3.6 \times 10^{-27}$ &  \\
6925  & 1.3 & 96  & $4.8 \pm 1.9$  & 81  & $2.8 \pm 1.4$ & $3.7 \times 10^{-12}$ & 6.9 \\
7500  & 1.4 & 75  &  $3.5 \pm 1.9$ & 78  & $2.3 \pm 0.9$ & $3.2 \times 10^{-4}$ & 3.6 
\enddata
\label{Table:stellartemps}
\tablenotetext{a}{The $p$-value and $n\sigma$ come from two-tailed KS tests run between the near-zero and random primary velocity simulations with the same stellar properties. $n\sigma$ refers to the level at which we can reject that the two groups are pulled from the same parent distribution. Below a $p$-value of $\sim 10^{-16}$, the confidence level that the populations are distinct approaches $\infty$. }
\end{deluxetable*}

These simulations reveal some interesting trends. They show that the near-zero primary velocity approach becomes even more beneficial when targeting hot Jupiters around late G-stars with lower temperatures, but less so for hot Jupiters around hotter stars. This could be due to a number of factors. Cooler stars have much more complex spectra which could allow for more improvement from well-chosen alignments of stellar and telluric features. They also have more spectral similarity with both their planets and the Earth. Water signatures arising in cooler stars could add to the need for carefully chosen, near-zero primary velocity epochs. Hotter stars, on the other hand, with fewer lines, will not be as affected by velocity shifts relative to the telluric frame.     

We also see quite a similarity between the heights derived from the random primary velocity cases across stellar temperatures and radii. This shows that there is a well balanced trade-off between more complex stellar spectra at lower stellar temperatures and lower planet/star contrast at higher stellar temperatures. 

These simulations indicate that the primary velocity trends observed in this work will be increasingly important to the study of hot Jupiters around cooler stars. We encourage future work into how more appropriate planet populations for each host stellar temperature and radius inform optimal high resolution, cross correlation observing strategies.

\section{Discussion}
\label{discuss}

\subsection{Applicability of Near-Zero $v_{pri}$ Observing Strategy}
We have seen that epochs during which the primary velocity is near zero, or, equivalently, the systemic velocity is canceled as much as possible by the barycentric velocity in the direction of the system so there is very little velocity separation between the telluric and stellar spectra, provide the strongest planetary detections. Not all systems will have periods during which the primary velocity is near zero however. The magnitude of the barycentric velocity, determined by a system's right ascension and declination, must be large enough to cancel out its systemic velocity. The second Gaia data release DR2 \citep{GaiaMission2016, Gaia2018} published radial velocities averaged over 22 months from 7,224,631 stars. These reported radial velocities were all from sources brighter than $G_{RVS} = 14$ (the flux measured in the Radial Velocity Spectrometer $G$ band); with a fraction of transits where the source was detect as having a double-lined spectrum less than 0.1 (to remove detected double-lined spectroscopic binaries); with an uncertainty on the radial velocity below 20 km/s; and a spectral template used to derive the radial velocity with an effective temperature from 3550 to 6900 K \citep{Gaia2018}. This was a substantial and collaborative effort and many researchers contributed to this impressive radial velocity data set \citep[e.g., ][]{Cropper2018, Sartoretti2018, Soubiran2018, Katz2019}. % describe the design and development of the Gaia Radial Velocity Spectrometer, \citet{Sartoretti2018} describe the Gaia spectroscopic data processing pipeline and the approach used to derive the radial velocities presented in DR2. \citet{Soubiran2018} compiled a catalogue of radial velocity standard stars to establish the radial velocity zero point provided in DR2, and \citet{Katz2019} describes the validation and properties of the median radial velocities published in DR2. 

Of the 7,224,631 stars with radial velocities reported in Gaia DR2, 3,209,212, or 44.4\%, have combinations of locations and systemic velocities that will allow for a near-zero primary velocity at some point during the year. This then suggests that our predicted optimal near-zero primary velocity strategy will then be applicable to nearly half of the planetary systems in the sky.

\subsection{Factors Influencing the Random Primary Velocity Distribution}
\begin{figure}
    \centering
    \noindent\includegraphics[width=21pc]{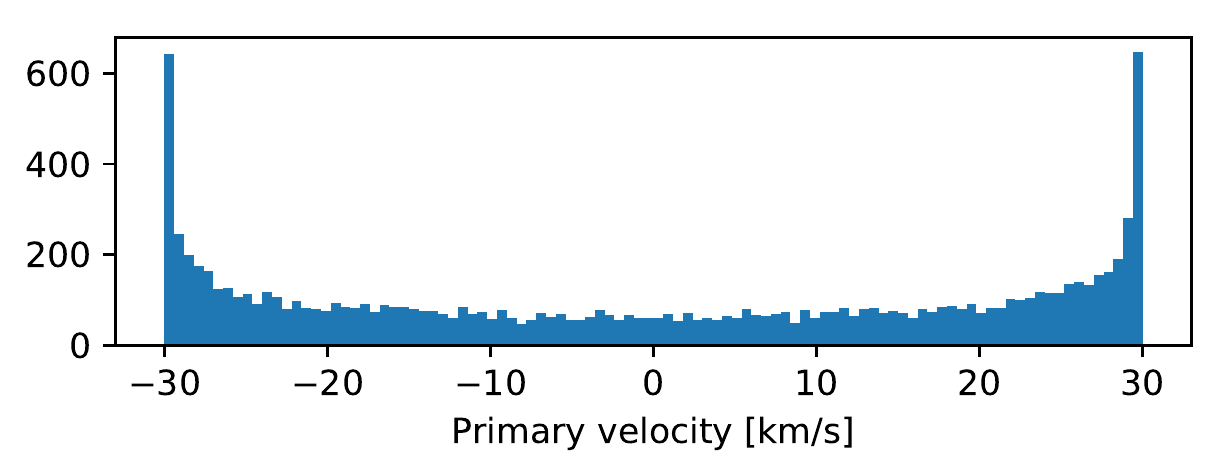}
    \caption{  ``Random" primary velocity distribution generated from 10,000 uniform pulls of Earth's orbital position.      } %nirspec1, ips from HD187123 5 nirspec 1 epochs, 1 jupiter radius, 1 solar radius }
    \label{Figure:randomvpriDistribution}
\end{figure}

In this work, to define the random primary velocity distribution, we pulled uniformly from the Earth's orbit, converted the Earth's position to a barycentric velocity assuming an orbital motion of 30 km/s, and then converted to primary velocity assuming a systemic velocity of 0. This resulted in a bimodal primary velocity distribution that was relatively flat through the center and rose quickly in probability toward $\pm30$ km/s (Figure~\ref{Figure:randomvpriDistribution}). It was this distribution from which we pulled ``random" primary velocities. 

This distribution could be different for different systems depending on their right ascensions, declinations, and systemic velocities, though. Their RAs and DECs will determine the component of the Earth's orbital motion that is in the line-of-sight to the system. While 30 km/s is about the Earth's actual orbital velocity, the barycentric velocity in the direction of a system would only vary from 30 to -30 km/s if the system were precisely on the Earth's orbital plane. Anywhere else and the range of possible barycentric velocities would shrink. A smaller range of primary velocities -- if the systemic velocity was still 0 km/s and so the primary velocity range still centered around 0 km/s -- might allow for slightly stronger detections than the random primary velocity simulations shown in this work. This would be because the random primary velocity distribution would have a smaller horizontal extent, so the values pulled would be nearer the optimal 0 km/s. 

Systems with non-zero systemic velocities would also have a differently shaped random primary velocity distribution than the one used in this work. The center of the distribution would now be around the systemic velocity, rather than zero, and the sharp increases in probability would be at the minimum ($v_{sys} - \mathrm{max}(v_{bary})$) and maximum ($v_{sys} - \mathrm{min}(v_{bary})$) ends of the distribution. We would expect systems with systemic velocities canceled by the maximum or minimum possible barycentric velocities along their line-of-sight to show the strongest detections possible with a random primary velocity sampling strategy. This would be because one of the sharp increases in towards the edges of the primary velocity distribution would be at 0 km/s, meaning that a random primary velocity strategy would offer the most near-zero primary velocity epochs if the system had this configuration of systemic and barycentric velocities (set by its RA and DEC) than any other alignment.  

Another consideration for the random primary velocity observing strategy will be the location of the telescope. The location of the telescope will set if and when during the year a system is observable, and so will cut out portions of the random primary velocity distribution corresponding to barycentric velocities that arise during times the system is not observable from that telescope. The primary velocities removed could include a range around zero or a range of the largest possible absolute primary velocities. If the primary velocities around 0 were removed, we would expect a weaker detection, while if a range of the largest absolute primary velocities were removed, we would expect a stronger detection.

We do note that these predictions are all based on which selection of random primary velocities would give the most near zero. We would still expect a dedicated near-zero primary velocity observing strategy to provide the strongest detection because it would not be diluted by any of the suboptimal non-zero primary velocity epochs that could arise from a random primary velocity sampling strategy regardless of systemic velocity, magnitude of the barycentric velocity variation, or telescope location. If the combination of the telescope location and systemic and barycentric velocities are such that there is no period of near-zero primary velocities, we would recommend targeting the smallest absolute (nearest zero) primary velocities as we saw provided stronger results in our pre-upgrade NIRSPEC simulations of Section~\ref{Section:ComparetoDataSection}.

\subsection{Random Orbital Phases}
In this work, we found that the combination of orbital phases did not have a large effect on the height over noise of the planetary detection. Our simulations considered cloud-free models and did not vary the planetary spectrum as a function of orbital phase, to account for day- to night-side differences for tidally locked planets, though. If day- to night-side differences were considered, we would expect the day-side orbital phases ($0.25 \leq M \leq 0.75$), which should have higher effective temperatures, to allow for stronger detections \citep{Finnerty}. 

While clouds have presented a challenge to low-resolution transmission spectroscopy, thermal emission spectra of the same planets show strong molecular lines \citep[e.g., ][]{Crouzet2014, Morley2017}. \citet{Gandhi2020} recently showed that high resolution transmission spectroscopy could be used to detect water and other trace species, namely CH$_4$, NH$_3$, and CO, in cloudy atmospheres with a modest observing time from a ground-based telescope. In high resolution emission spectra, clouds could decrease the line contrast by shifting the continuum to higher altitudes and lower temperatures, rather than by blocking stellar rays below the cloud tops as they do in transmission spectra. By decreasing line contrasts, clouds would make the planet more difficult to detect through cross correlation analysis. If, in tidally-locked atmospheres, the clouds are mainly constrained to the night-side \citep[e.g., ][]{Demory2013, Parmentier2016}, day-side epochs would be even more preferable.  

For longer period planets that are not tidally-locked, neither day- to night-side temperature differences nor night-side clouds would uniformly degrade one set of orbital phases over another. Additionally, neither day- to night-side differences nor the presence of clouds should affect our predictions for the optimal primary velocity observing strategy. 

Importantly, the fact that random orbital phases are sufficent, at least for non-tidally locked atmospheres, indicates that a robust detection could be made with only a fraction of an exoplanet's orbital period. Short period planets could be well detected with a selection of orbital phase epochs taken over a period when $v_{pri}$ is near zero. With the much more quickly varying planetary orbital phase relative to Earth's orbital phase, these periods during which $v_{pri}$ is near zero should offer a range of day- to night-side planetary epochs. Longer period planets could be targeted at multiple stretches when $v_{pri}$ is near zero, each offering a different selection of orbital phases (as long as the orbital period is not highly commensurate with that of the Earth). Such observing strategies could be easily obtainable and should lead to strong (non-transiting) planetary detections.

\subsection{Wavelength Dependence and Atmospheric Characterization}
Further, while we investigated ways to strengthen detections of planetary emission through the recovery of $K_p$ in this work, ultimately, we would be interested in constraining various planetary atmospheric properties, such as the presence and relative abundances of various molecular species and the natures of the atmospheric thermal structure, winds, and planetary rotation. Previous work has found that since there are no spectral lines from major carbon-bearing species in the $L$ band of hot Jupiter atmospheres, this data alone is not sufficient to constrain their atmospheric C/O ratios \citep[e.g., ][]{Piskorz2018,Finnerty}. Such measurements may be possible for warm Jupiters ($T_{\mathrm{eff}} \approx 900$ K) from $L$ band data alone however. At cooler effective temperatures, sufficient methane can be expected under equilibrium conditions to be detectable in $L$ band data. With both methane and water appearing, $L$ band data can provide constraints on the C/O ratios of warm Jupiters \citep{Finnerty}. 

To make these C/O constraints for hot Jupiters would likely require additional epochs in the $K$ or $M$ bands, where prominent carbon monoxide bandheads exist. In this work, we found that both pre- and post-upgrade NIRSPEC $L$ band simulations were better able to detect planetary signals with near-zero, rather than with random, primary velocity epochs. The pre- and post-upgrade simulations differ in both number of orders per epoch and order wavelength coverage, with no overlap between the wavelengths covered. The fact that both still preferred a near-zero primary velocity epoch strategy implies that these predictions are not completely wavelength dependent and we expect that they should hold for $L$ band observations in general. We encourage more simulation work to determine how widely generalizable these predictions will be both at other NIRSPEC bands (specifically $K$ and $M$) and across the large instantaneous spectral grasp promised by upcoming and proposed instruments such as GMTNIRS (1.1--5.3 $\mu$m) and IGNIS (1--5 $\mu$m). Data covering these multiple bands would allow us to detect carbon monoxide as well as water in hot Jupiter atmospheres and allow for constraints on their atmospheric C/O ratios.

\section{Conclusion}
\label{conclude}

In this work, we aimed to determine how to best strengthen planetary detections and reduce structured noise in few epoch data sets with careful observing strategies. The two key parameters that can be selected with the choice of observing nights are the primary velocity (because of the variable barycentric velocity) and the planetary orbital phase. We found that epochs taken during nights when the primary velocity of the system is near 0 km/s, so that there is very little relative velocity shifting of the stellar and telluric reference frames, will provide the strongest planetary detections. With a random selection of planetary orbital phases, these near-zero primary velocity epoch simulations produce planetary peaks more than two times higher relative to the noise than simulations generated with randomly selected primary velocities. Further, for near-zero primary velocity epochs, the closer their orbital phases are to quadrature, the better the constraints on $K_p$ will be. Following these results, we recommend that observers looking to build up multi-epoch near-IR high resolution data sets target, first, epochs with near-zero primary velocities, and second, epochs with orbital phases near quadrature to get the best constraints on the planetary detection. In this work, we demonstrated how greatly the combinations of primary velocities and orbital phases can affect a planetary detection. Moving forward, careful attention should be paid to planning observations, and all few epoch data sets should not be assumed to have an equal probability of detecting a planet. Following these predications, observations taken from upcoming multi-echelle instruments, such as GMTNIRS and IGNIS, during periods when the primary velocity of a system is near zero could provide both robust detections of exoplanets in a fraction of their orbital periods and constraints on their atmospheric composition.

\acknowledgments{

The authors wish to recognize and acknowledge the very significant cultural role and reverence that the summit of Mauna Kea has always had within the indigenous Hawaiian community. We are most fortunate to have the opportunity to conduct observations from this mountain. The data presented herein were obtained at the W. M. Keck Observatory, which is operated as a scientific partnership among the California Institute of Technology, the University of California and the National Aeronautics and Space Administration. The Observatory was made possible by the generous financial support of the W. M. Keck Foundation. 

This work has made use of data from the European Space Agency (ESA) mission
{\it Gaia} (\url{https://www.cosmos.esa.int/gaia}), processed by the {\it Gaia}
Data Processing and Analysis Consortium (DPAC,
\url{https://www.cosmos.esa.int/web/gaia/dpac/consortium}). Funding for the DPAC
has been provided by national institutions, in particular the institutions
participating in the {\it Gaia} Multilateral Agreement. 

We would also like to thank an anonymous reviewer who pointed us in interesting new directions and by doing so improved the content of this paper. }

\appendix

\subsection{Notes on Epochs from Individual Sources} \label{Section:appendixnotes}

All of the data used in Section~\ref{Section:ComparetoDataSection} is $L$ band data from the pre-upgrade NIRSPEC instrument. Each epoch has 4 orders, covering approximately 2.9962--3.0427, 3.1203--3.1687, 3.2552--3.3058, and 3.4026--3.4554 $\mu$m. The average spectral resolution is 20,000, and the total S/N across the 5 epochs is about 4100.  

\subsubsection{HD187123}
The reduced data and PHOENIX stellar and SCARLET planetary spectral models used here were those presented in \citet{Buzard2020}.

\subsubsection{KELT2A}
 The reduced data and PHOENIX stellar and planetary models used here were those presented in \citet{Piskorz2018}. For the planetary model, we used the best fitting ScCHIMERA model, which had a metallicity ($\log z$) of 1.5, a C/O ratio of 0.5, and an incident solar flux $f$ of 1.0. This parameter $f$ accounts for day-night heat transport and an unknown albedo by scaling a wavelength-dependent incident stellar flux (from a PHOENIX stellar grid model). Defined this way, model atmospheres with $f \gtrsim 1.5$ show a temperature inversion.

\subsubsection{51 Peg}
The 51 Peg epochs were reduced in the same way as the other epochs \citep[e.g., ][]{Piskorz2018, Buzard2020}, and telluric corrected through a Molecfit \citep{Kausch2014} guided principal component analysis. We used a PHOENIX stellar spectral model interpolated to an effective temperature of 5787 K, a metallicity of 0.2, and a surface gravity of 4.449 \citep{Turnbull2015}. The planetary spectral model we use was generated from the SCARLET framework. It does not have an inverted thermal structure, which was suggested is appropriate by \citet{Birkby2017}.

\subsubsection{Tau Boo}
The Tau Boo data used here were processed using a Molecfit initial telluric model followed by PCA to remove residual tellurics. The stellar and planetary spectral model used here were the ones used in \citet{lockwood}. The stellar model was not from the PHOENIX framework. Rather, it was generated from the LTE line analysis code MOOG \citep{Sneden1973} and the MARCS grid of stellar atmospheres \citep{Gustafsson2008}. Individual elemental abundances were set through fitting to well measured lines in the NIRSPEC data. See \citet{lockwood} for a full description of the stellar spectral model generation.

\begin{figure*}
    \centering
    \noindent\includegraphics[width=42pc]{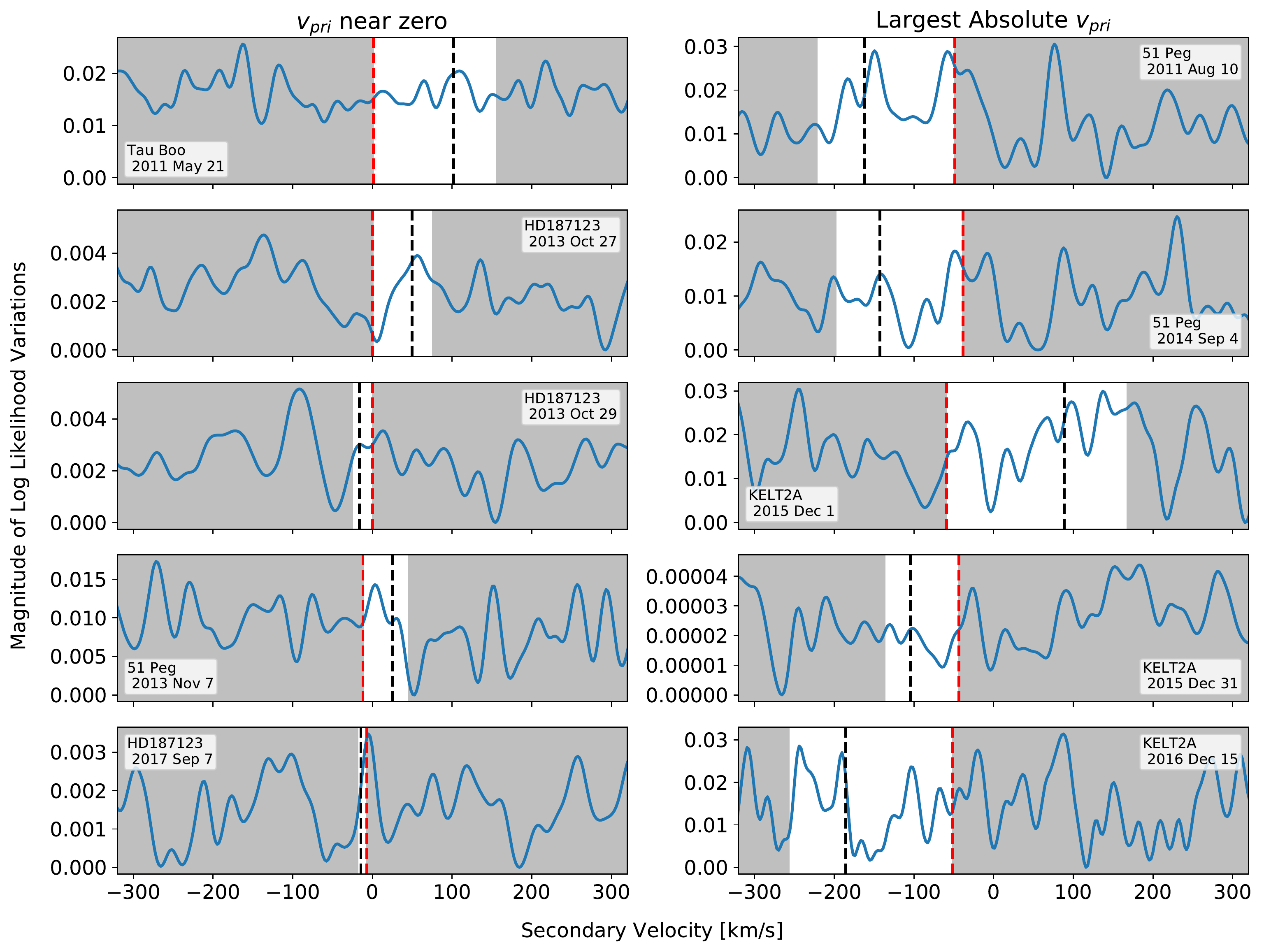}
    \caption{ Magnitude of the log likelihood variations of each epoch in the two data primary velocity groups -- near-zero and largest absolute primary velocity. Each curve has been converted to reflect a $K_p$ of 150 km/s, rather than the underlying planets' true $K_p$ values, which are reported in Table~\ref{targetstable}. These curves, converted to $K_p$ space and summed, make up Figure~\ref{DataFigure}. In each subplot, the red dashed line corresponds to the primary velocity at that epoch and the black dashed line corresponds to the $v_{sec}$ given by a $K_p'$ of 150 km/s at that each. If the fictitious combined system were face-on, with a $K_p$ of 0 km/s, the black dashed line would coincide with the red dashed line. If, on the other hand, it were edge-on, $v_{sec}$ would fall on the other end of the white range of possible planetary velocities. Here we have the maximum value of $K_p$ arbitrarily set to 230 km/s.     }
    \label{mlcombinedatafigure}
\end{figure*}

{\footnotesize
\bibliography{sample}}
\bibliographystyle{ApJ}

\end{document}